\documentclass[twocolumn, 10pt, aps, pra]{revtex4-1}

\usepackage{graphicx}
\usepackage{amsmath}
\usepackage{amssymb}
\usepackage[usenames, dvipsnames]{xcolor}
\usepackage{appendix}
\usepackage[colorlinks=true, pdfstartview=FitV, linkcolor=blue, citecolor=blue, urlcolor=blue]{hyperref}
\usepackage{color}
\usepackage{epstopdf}
\usepackage{dsfont} 
\usepackage{bm,bbm}

\usepackage[super]{nth}

\DeclareMathOperator{\tr}{tr}

\begin{document}

\title{	
	Quantum adiabatic and diabatic optimization: \\ A Hilbert space engineering strategy with Rydberg arrays
}
 
\author{Rui Lin}
\email{linrui@quantumsc.cn}
\affiliation{Quantum Science Center of Guangdong-Hong Kong-Macao Greater Bay Area, 518045 Shenzhen, China}

\date{\today}

 
\begin{abstract}
Neutral-atom arrays provide a versatile platform for encoding combinatorial optimization problems with arbitrary graph connectivity [Nguyen et al., PRX Quantum 4, 010316 (2023)]. Although the dynamics of individual encoding gadgets has been studied [Bombieri et al., PRX Quantum 6, 020306 (2025)], understanding annealing bottlenecks in larger networks remains challenging. We investigate how engineering the Hilbert-space geometry controls quantum annealing in a family of blockaded-chain models motivated by these encodings. While boundary terms encode the target maximum independent set problem, the bulk geometry induces an auxiliary optimization problem associated with independent sets of cliques. Spectrally, the energetics of the quantum system can be approximated by physically derived clique weights. When the target problem and the auxiliary problems are consistent, the low-energy modes lie predominantly within the target subspace. This spectral structure motivates a diabatic annealing strategy that exploits target-compatible excited states, as tested on the diamond-chain graph family. These results provide a framework for predicting annealing bottlenecks, and demonstrate Hilbert-space engineering as a strategy for improving neutral-atom quantum optimization protocols.

\end{abstract}
 
\maketitle

\section{Introduction}

Quantum many-body systems can be exploited to solve combinatorial optimization problems. Within these encoding schemes, the optimal solution to the classical problem is mapped to the ground state of the quantum many-body system. 
One method to search for the ground state of a quantum many-body Hamiltonian in a quantum platform is through quantum annealing algorithm~\cite{johnson11quantum,kim25quantum}. The system is initialized in a known and easy-to-prepare state, and its parameters are slowly ramped toward the target. In the adiabatic limit, the state follows the instantaneous ground state. Revealing the source of bottleneck encountered in quantum annealing, and subsequently improving the annealing performance typically requires a comprehensive and oftentimes intractable study on the quantum many-body dynamics. 
Conceptually, departure from the instantaneous ground state need not imply failure to solve the encoded optimization problem. It is thus crucial to identify and utilize the compatibility of excitations with the target solution in the encoding.

Recently, neutral atom arrays have become a promising platform for quantum computing and quantum simulations, due to their rapid and substantial development in theoretical proposals and experimental techniques~\cite{saffman10quantum,labuhn16tunable,ebadi21quantum,cesa23universal,evered23high,grun24optical,norcia24iterative,pichard24rearrangement,cheng24emergent,zhang24entanglement,manetsch25tweezer,lin25AI,chiu25continuous,mohan25parametrized,qiao25realization,qiao2025kinetically,zhang25observation,wang26direct,senoo26high,zeng26adiabatic,cesa26engineering,white26quantum,bluvstein26fault}.
The Rydberg blockade mechanism~\cite{urban09observation,browaeys20many} of neutral atoms provides a natural quantum encoding scheme for a family of combinatorial optimization problems~\cite{pichler2018quantum,kim22rydberg,byun22finding,ebadi22quantum,nguyen23quantum,oliveira25demonstration,schuetz25quantum,bombieri25quantum,cazals26identifying,deoliveira2026quantum,kombe26reducibility}, including the unweighted maximum independent set problems.
The encoding scheme for these problems has been investigated theoretically~\cite{schuetz25quantum,cazals26identifying,kombe26reducibility,byun26error} and experimentally~\cite{ebadi22quantum,kim22rydberg,byun22finding,oliveira25demonstration,deoliveira2026quantum}.
Recently, a strongly correlated and geometrically localized many-body bound state has been shown to be the major source of annealing bottlenecks within a building block of the network, called \emph{gadget}. Correspondingly, a strategy controlling the effective shape and boundary conditions of the system has been proposed to alleviate the difficulty ~\cite{bombieri25quantum}. However, the effects of these dynamical bottleneck in more complex networks is yet to be unveiled.

In this work, we elucidate the low-energy dynamics of the Rydberg networks using the effective blockaded-chain models. While the boundary condition encodes the target optimization problem by design, the bulk of the system naturally encodes a pertinent auxiliary optimization problem. Specifically, quantum many-body bound states contribute characteristic energies, allowing competing many-body configurations to be organized through weighted cliques of the graph. In the quantum system, this corresponds to energetic competition between quantum states lying in macroscopically distinct Hilbert subspaces, and leads to bottleneck in quantum annealing.

Compatibility between the target and auxiliary problems can favor low-energy states with substantial support in the target subspace. This spectral structure suggests that excitation out of the instantaneous ground state need not prevent eventual recovery of the desired target readout. We investigate this possibility through a diabatic annealing protocol on diamond-chain graphs, relating the observed dynamics to competition between target-compatible and suboptimal channels.

This work is structured as follows. 
Section~\ref{sec:encoding_hamiltonian} introduces the encoding, effective model, and performance diagnostics. Sections \ref{sec:hypercube}, \ref{sec:product_state}, \ref{sec:analysis} develop the spectral description, and investigate how geometric constraints and quantum many-body effects like hybridization and binding lead to clique-based quantum states, and how different branches compete. Section~\ref{sec:dynamics} examines the resulting dynamics on diamond-chain graphs, followed by implementation considerations and conclusions in Sec.~\ref{sec:conclusions}.

\section{Encoding scheme and Hamiltonian}
\label{sec:encoding_hamiltonian}

\subsection{Introduction to the graph theory problems}
\label{sec:background}

\begin{figure*}[!t]
	\centering
	\includegraphics[width=\textwidth]{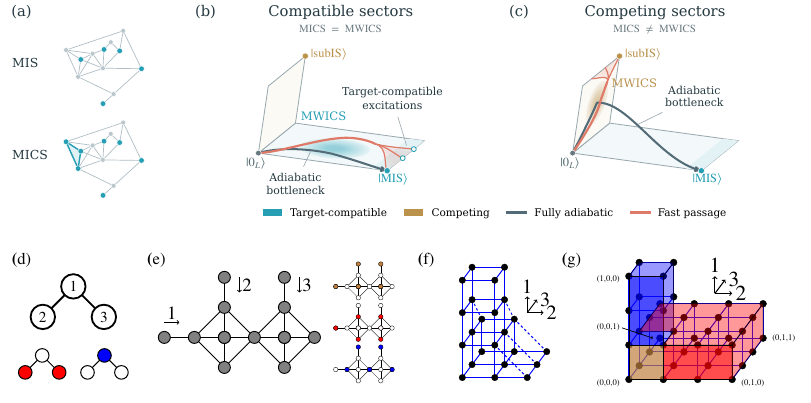}
	\caption{
		{\bf Graph theoretical representation of quantum many-body states and dynamics for quantum annealing.}
		(a) Maximum independent vertex and clique sets (MIS, MICS) of a graph. The two sets are consistent: choosing a vertex from each clique in MICS yields the MIS.
		(b,c) Schematics associating quantum annealing bottleneck to features of the graph. Along the annealing, the ground state behaves like boundary states representing the logical-zeros state $|0_L\rangle$ and the target $|\mathrm{MIS}\rangle$ state in the initial and final stages, respectively. In the intermediate stage, the ground state spreads in the bulk of the system, encoding the solution to the maximum weight independent clique set (MWICS) problem. This solution can be (b) consistent with the MICS problem or (c) not, depending on the graph and the system parameters. Following the ground state (solid arrows) requires traversing narrow gaps and costs long time overheads.
		(b) Particularly, when MICS and MWICS are consistent, the low-energy quantum many-body spectrum allows a diabatic annealing scheme for reaching the target MIS state while bypassing the bottleneck.
		(d) The $N=3$ chain graph and its (red) maximum and (blue) suboptimal independent sets.
		(e) A realistic implementation using neutral atom arrays subject to Rydberg blockade mechanism for encoding the MIS problem. The filled circles are atoms, and the links between them indicate Rydberg blockade. Three configurations are shown representing (brown) logical-zero, (red) optimal solution, and (blue) suboptimal solution, respectively, in the classical limit. The Rydberg mode (ground mode) of the neutral atoms is occupied when the atom is colored (white). 
		(f) The effective Hamiltonian describing the domain wall states in the low energy sector of the neutral atomic system shown in (e), which motivates our blockaded lattice model. Particularly, the dashed lines highlight the hoppings which are explicitly omitted in our blockaded lattice model. 
		(g) Schematics illustrating the blockaded lattice model encoding the MIS problem on the $N=3$ chain graph $\mathcal{G}_{3-1}=\mathcal{P}_3$, with $L=4,d=2$. 
	}
	\label{fig:illustration}
\end{figure*}

The \emph{maximum independent set} (MIS) problem in graph theory provides a stage for examining quantum computing algorithms. A connected simple graph $\mathcal{G}=(\mathcal{V},\mathcal{E})$ consists of a vertex set $\mathcal{V}$ and an edge set $\mathcal{E}$ connecting the vertices. Given a graph, a subset of vertices is said to be independent, when no pair of vertices share an edge. The MIS problem thus searches for the independent set with the largest cardinality, see Fig.~\ref{fig:illustration}(a) for an example. This set, represented by the array $\sigma^\mathrm{MIS}$, minimizes the loss function
\begin{eqnarray}\label{eq:mivs_constraint}
	H = -\sigma^T \sigma, \quad \text{subject to}\quad   \sigma^T \mathcal{A} \sigma = 0,
\end{eqnarray}
Here, $\sigma=(\sigma_1,\dots,\sigma_N)^T$ with $\sigma_n=0,1$ denoting the occupancy of vertex $n$, and $N=|\mathcal{V}|$ is the cardinality of the graph. $\mathcal{A}$ is the adjacency matrix encoding the edges, $\mathcal{A}_{m,n}=0$ for $<m,n> \notin \mathcal{E}$ and $\mathcal{A}_{m,n}=1$ for $<m,n> \in \mathcal{E}$. Generically, the MIS problem is NP-complete, and an exact and generic classical branching algorithm has a  complexity scaling exponentially with the size of the graph $O(\gamma^{N})$, with $\gamma\approx1.26$~\cite{gao2024automated}. 

Neutral atoms are subject to the Rydberg blockade mechanism, which can be effectively described by the PXP model in the hardcore limit~\cite{jaksch00fast,lukin01dipole,gaetan09observation,serbyn21quantum,su23observation}. The Rydberg states of two spatially adjacent atoms cannot be occupied simultaneously. This constraint on the quantum device is naturally reminiscent of the graph-theoretical constraint Eq.~\eqref{eq:mivs_constraint}. This correspondence motivates a proposal for encoding the MIS problem on unit-disk graphs using arrays of neutral atoms~\cite{pichler2018quantum,ebadi22quantum,cain2023quantum}. Later, the mapping proposed in Ref.~\cite{nguyen23quantum} generalizes the encoding scheme to graphs with arbitrary connectivity, and eventually motivates the model investigated in this work. Once quantum encoded, the solutions to the classical optimization problem can be read out from the ground states, which are to be solved by \emph{quantum annealing algorithms}~\cite{johnson11quantum,kim25quantum}. The performance is determined by the low-energy dynamics of the instantaneous Hamiltonians along the entire annealing process.

In this work, we find that the quantum many-body dynamics can be inferred from a classical indicator, termed maximum weight independent clique set (MWICS), as illustrated in Fig.~\ref{fig:illustration}(a).
A clique $K\subseteq\mathcal{G}$ is a fully connected subgraph of cardinality $q=|K|=\sum_{n=1}^{N}\sigma^K_n$, satisfying $(\sigma^{K})^{T}\mathcal{A}\sigma^K=q(q-1)$. Here, the clique is represented by an array $\sigma^K$ listing a set of vertices $\mathcal{S}$ inducing the clique, $K = \mathcal{G}[\mathcal{S}]$.
A set of cliques, $\mathcal{K}=\{K_1,K_2,\dots K_R\}$, is said to be independent, when vertices from these disjoint cliques are not connected by any edge, i.e., $(\sigma^{K}_r)^{T} \sigma^K_s =0$, and $(\sigma^{K}_r)^{T}\mathcal{A} \sigma^K_s =0,r\neq s$. 
The MWICS problem associates each clique with a weight $W_{r}$ depending on its cardinality, and search for the independent clique set (ICS) with the largest weight. It minimizes the loss function
\begin{eqnarray}\label{eq:mics_constraint}
	&&H = -\sum_{r=1}^{R} \, W_{r}, \\
	 \text{subject to}\quad  &&(\sigma^{K}_r)^{T}\mathcal{A} \sigma^K_s = (q_r^2-q_r) \delta_{r,s}. \nonumber
\end{eqnarray}
In the same spirit, MWICS is reduced to the maximum independent clique set (MICS) problem, by assigning weights $W_{r} = \lim_{\varepsilon\to0^+}1+q_r\varepsilon$, see Fig.~\ref{fig:illustration}(a) for an example.  MIS and MICS are naturally compatible with each other, and a MIS can be straightforwardly read out from a MICS. In this work, we will show that the annealing difficulty can be alleviated when MWICS is reduced to the MICS, see Fig.~\ref{fig:illustration}(b,c).

\subsection{The blockaded chain Hamiltonian}

We encode the MIS problem in the \emph{blockaded chain model},
\begin{eqnarray}\label{eq:model}
	H_{(L,d,\mathcal{G})} = \Pi_{(L,d,\mathcal{G})} H_{L}^N \Pi_{(L,d,\mathcal{G})}
\end{eqnarray}
Here, the bare Hamiltonian $H_{L}^N$ describes $N$ independent chains with nearest-neighbor hopping strength $t$ and local chemical potentials $\epsilon_{i,f}$ at the two edges.
\begin{subequations}
	\label{eq:intact_cube}
	\begin{eqnarray}
		H_{L}^N &=& \sum_{n=1}^{N=|\mathcal{V}|} h_{L,n} \\
		h_{L,n} &=&  -t \sum_{\ell=1}^{L-1} \left(|n,\ell\rangle \langle n,\ell+1| + \mathrm{h.c.}\right) \nonumber \\
		&&- \epsilon_i |n,1 \rangle\langle n,1| 
		 - \epsilon_f |n,L \rangle\langle n,L| 
		\label{eq:intact_cube_onebody}
	\end{eqnarray}
\end{subequations}
This bare Hamiltonian describes $N$ non-interacting chains of length $L$, with lattice sites denoted as $|n,\ell\rangle,n=1,\dots,N,\ell=1,\dots,L$. Notably, it can also be interpreted as an $N$-dimensional lattice in the shape of a hypercube, and encodes the MIS problem for a graph consisting of $N$ unconnected vertices. The projection operator $P$ encodes the independent set constraint associated to the graph, by projecting out the relevant lattice sites from being occupied,
\begin{subequations}
	\label{eq:projection_operator}
	\begin{eqnarray}
		\Pi_{(L,d,\mathcal{G})} &=& \prod_{<m,n>\in \mathcal{E}} \Pi_{<m,n>} \\
		\Pi_{<m,n>}  &=&  \mathds{1} - \sum_{\ell_m,\ell_n=d+1}^{L} |m,\ell_m\rangle |n,\ell_n\rangle \langle m,\ell_m| \langle n,\ell_n | \nonumber \\
	\end{eqnarray}
\end{subequations}
Particularly, $<m,n>$ refers to an edge between vertices $m$ and $n$ in the graph, and the projection operation can be viewed as an interaction between the $m$th and $n$th chains of infinite strength.

Given a specific graph, $L$ and $d$ are the two important control parameters. In this work, we refer to $d=L-1,L\neq2$ as the hypercube limit, $L\gg d$ as the indent limit, and $L=2,d=1$ as the \emph{PXP implementation}. Particularly, for $L\gg1$, the system can be treated as a continuum, and the ratio $d/L\in (0,1)$ solely determines the shape of the Hilbert space. As an example, the lattice hosting $H_{(L,d,\mathcal{P}_3)}$ is shown in Fig.~\ref{fig:illustration}(g). Notably, $H_{(L=2,d=1,\mathcal{P}_N)}$, with $\mathcal{P}_N$ being chain graphs of $N$ vertices, is the commonly discussed PXP model~\cite{bluvstein21controlling,hudomal22driving,giudici24unraveling,park25nonintegrability}.

We refer to $H_{(L,d,\mathcal{G})}$ as the $(L,d,\mathcal{G})$-blockaded lattice model. 
It is motivated by experimentally relevant neutral atom arrays. A scheme for encoding the MIS problem on a general graph has been proposed in Ref.~\cite{nguyen23quantum}, which utilizes the Rydberg blockade mechanism of neutral atoms for naturally encoding the independent-set constraint on a quantum device. The crucial building blocks are the \emph{atom wires} representing the graph vertices, and the \emph{gadgets} lying at the wire crossings representing the connectivity between the two wires. The blockaded-chain model effectively captures the low-energy physics of these Rydberg networks, see Appendix~\ref{sec:chain_from_network}.

The quantum dynamics within these building blocks has been thoroughly investigated in Ref.~\cite{bombieri25quantum}. Particularly, the low energy quantum dynamics of the system has been found to be governed by domain walls along the atom wires. For a two-vertex connected  graph $\mathcal{G}_{2,1}=\mathcal{K}_2$, the domain walls travel on a two-dimensional lattice, where part of the lattice sites are prohibited, forming an $\mathsf{L}$-shaped lattice. A similar treatment for the graph $\mathcal{G}_{3,1}=\mathcal{P}_3$ leads to the Rydberg array illustrated in Fig.~\ref{fig:illustration}(e), and the effective lattice model shown in Fig.~\ref{fig:illustration}(f). For simplicity, we impose that the gadget is placed at the same position along each wires, leading to a symmetric implementation. Experimentally, this corresponds to elongating the every wire along both directions at the same proportionality, making the size and detailed structures of the gadgets negligible in comparison to the entire wires. Moreover, hopping terms at the gadgets linking next-nearest-neighbor lattice sites are also omitted. These extra hopping terms have been shown not to exhibit essential contributions to quantum dynamics~\cite{bombieri25quantum}. Taking all these considerations into account, the $H_{(L,d,\mathcal{K}_2)}$ and $H_{(L,d,\mathcal{P}_3)}$ are achieved, respectively. In Ref.~\cite{bombieri25quantum}, geometric confinement has been found as a substantial source of obstacle for a polynomial scaling of the annealing efficiency with respect to $L$ in the $\mathsf{L}$-shaped domain associated to the two-vertex graph, $\mathcal{K}_2$. Nevertheless, the quantum annealing performance associated to $H_{(L,d,\mathcal{G})}$ for a more general graph $\mathcal{G}$ is so far not yet explored.

\subsection{Hilbert space}
\label{sec:hilbert_space}

The blockaded chain model provides a suitable platform for solving the MIS problem in two perspectives. Straightforwardly, the ground states under selected boundary conditions are connected to logical configurations of the graph. When $\epsilon_i\gg t,\epsilon_f=0$, the instantaneous ground state can be interpreted as logical-zero state,
\begin{subequations}
	\label{eq:initial_target_states}
	\begin{eqnarray}
		|0_L\rangle = \bigotimes_{n=1}^{N}|n,1\rangle;
	\end{eqnarray}
	while when $\epsilon_i=0,\epsilon_f\gg t$, the quantum state is localized in the vicinity of the lattice site encoding the MIS solution $\{\sigma^\mathrm{MIS}_n\}$,
	\begin{eqnarray}
		|\Psi^\mathrm{MIS}\rangle = \bigotimes_{n=1}^{N}|n,\sigma^\mathrm{MIS}_n\times (L-1)+1\rangle.
	\end{eqnarray}
\end{subequations}

Nevertheless, the blockaded chain model encodes the optimization problem beyond its boundary, and also in its bulk. The compatibility between the boundary and the bulk fundamentally facilitates the quantum annealing processes. To gain a first insight, we inspect the structure of the intact $N$-dimensional hypercube lattice $\bigotimes_{n=1}^{N}\bigoplus_{\ell=1}^{L} |n,\ell\rangle\langle n,\ell|$ projected out by $\Pi_{L,d,\mathcal{G}}$ in Eq.~\eqref{eq:projection_operator} from a constructive point of view~\cite{wu20quantum,zhao25quantum},
\begin{eqnarray}
	\label{eq:hilbert_space_structure}
	&& \Pi_{L,d,\mathcal{G}} \bigotimes_{n=1}^{N}\bigoplus_{\ell=1}^{L} |n,\ell\rangle\langle n,\ell| \Pi_{L,d,\mathcal{G}} \nonumber \\
	&=& {\bigcup_{\sigma\in \{\sigma^\mathrm{mIS}\}}} \left(\bigotimes_{n=1}^{N} \bigoplus_{\ell=1}^{\ell_n^{L,d}}|n,\ell\rangle\langle n,\ell|\right)
\end{eqnarray}
The construction is directly associated to the collection of \emph{maximal} independent sets (mIS), $\{\sigma^\mathrm{mIS}\}$, of the graph $\mathcal{G}$. They are ISs which are not a proper subset of another IS.
Given an mIS, $\bigotimes_{n=1}^{N} \bigoplus_{\ell=1}^{\ell_n^{\mathrm{mIS},L,d}}|n,\ell\rangle\langle n,\ell|$ constructs an $N$-dimensional hypercuboid, and its lengths along dimension $n$ is given by $\ell^{L,d}_n=d(L)$ for $\sigma_n=0(1)$. 
The projected Hilbert space can be constructed by merging hypercuboids of different mISs. The structure of the lattice mimics a Daisy cube~\cite{klavzar2017daisy,taranenko20daisy} generated by all mISs of the graph, particularly for the PXP implementation $L=2,d=1$.

The structure of the Hilbert space of the blockaded chains alludes to quantum dynamics. A hypercuboid Hilbert subspace acts locally like an integrable system, and admits tractable quantum dynamics. Moreover, the MIS, being the classical optimal solution among all mISs, corresponds to the hypercuboid of the largest volume in this construction. Intuitively, this strongly facilitates the propagation of the quantum state and the annealing process for searching the MIS solution. This effect should be most prominent in the indent limit $L\gg d$.

\subsection{Logical annealing protocol and performance}
\label{sec:protocol}

\begin{figure}[!t]
	\centering
	\includegraphics[width=\columnwidth]{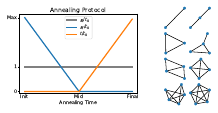}
	\caption{
		{\bf  Quantum annealing protocol and list of graphs.} (a) The annealing parameters, $t$, $\epsilon_{i}$, and $\epsilon_{f}$, as a function of annealing time in the logical protocol. (b) A list of small graphs used as examples for discussions in this work. An exhaustive atlas of small graphs with $N\leq 5$ vertices is listed in Appendix~\ref{sec:graph_list}.
	}
	\label{fig:protocol}
\end{figure}

The objective of the quantum annealing is to adiabatically evolve the quantum many-body state from the initial state to the target state in an unbiased manner. The unbiasedness requires that (i) the initial state is known and easy to prepare, which is fulfilled by Eq.~\eqref{eq:initial_target_states}; (ii) the structure and the parameters of the Hamiltonian is not biased toward/against certain solutions, e.g., a sequential ramping of boundary terms associated different vertices is prohibited; (iii) the annealing protocol should not be based on learned, non-universal features associated to a specific graph, e.g.\ slowing down the annealing for addressing a fine structure of the spectrum of a specific Hamiltonian is prohibited. 

The \emph{logical annealing protocol} proposed in Ref.~\cite{bombieri25quantum} and shown in Fig.~\ref{fig:protocol} meets these criteria. It further provides a tractable arena for investigating the quantum dynamics within the blockaded chain model. In this protocol, the initial boundary chemical potential $\epsilon_i$ is first linearly turned off, and subsequently the final boundary chemical potential $\epsilon_f$ is linearly turned on. Particularly, the boundary terms of every chain are ramped simultaneously. Meanwhile, the hopping $t$ is kept fixed during the whole process, cf.\ Eq.~\eqref{eq:intact_cube}. The logical protocol can be straightforwardly implemented upon the atom network, and allows for an intuitive analysis of quantum dynamics, see Appendix~\ref{sec:chain_from_network}.

We characterize quantum annealing performance using complementary spectral and dynamical indicators. 
Spectrally, quantum adiabatic theorem~\cite{schaller06general,jansen07bounds,boixo10necessary} measures the performance using the minimum gap of the instantaneous Hamiltonians within the dynamically accessible branches. This is most clearly seen when the solution to the MIS problem is unique. In this case, due to the uniqueness of ground state, the annealing performance can be represented by the minimum gap between the instantaneous ground first excited state along the annealing, $\Delta_\mathrm{min} = \min(E_1 - E_0)$. However, when the MIS problem has $k\ge2$ degenerate solutions, there is no a general rule for assigning a gap for quantifying the performance, and we tentatively employ $\Delta_\mathrm{min} = \min(E_k - E_0)$ for this purpose. In this work, the spectrum is solved numerically using \emph{exact diagonalization} of sparse matrices unless specified otherwise.

Dynamically, the quantum annealing performance can be measured by the success rate of the quantum state reaching the target state, or more generically, the target subspace. We propose the dephased density probability in the MIS subspace,
\begin{eqnarray}
	\label{eq:def_PT}
	P_T = \sum_\mu |\langle\mu|\Psi\rangle|^2 \langle \mu | \Pi_{\mathrm{MIS}} | \mu\rangle,
\end{eqnarray}
where $|\mu\rangle$ are the instantaneous eigenmodes. $\Pi_{\mathrm{MIS}}= \bigotimes_{n\in\mathcal{U}}\Pi_n^{(0)}$ projects onto (the union of) the logical hypercuboids associated with the MIS (degenerate MISs). Here, $\mathcal{U}$ contains all vertices that are not in (any of the degenerate) MISs, and $\Pi_n^{(0)}=\sum_{\ell=1}^{d}|\ell\rangle_n\langle\ell|$. Notably, $P_T$ measures the weight of the quantum state in the entire MIS subspace, rather than its fidelity to a particular state. It further averages out the phases between different $|\mu\rangle$ to account for an average in time in dynamics. Complementary and experimentally relevant information can be further provided by marginal density distribution along a dimension associated to logical-one, see Appendix~\ref{sec:chain_from_network}.

\section{Quantum annealing performance in hypercube limit}
\label{sec:hypercube}

\begin{figure}[!t]
	\centering
	\includegraphics[width=\columnwidth]{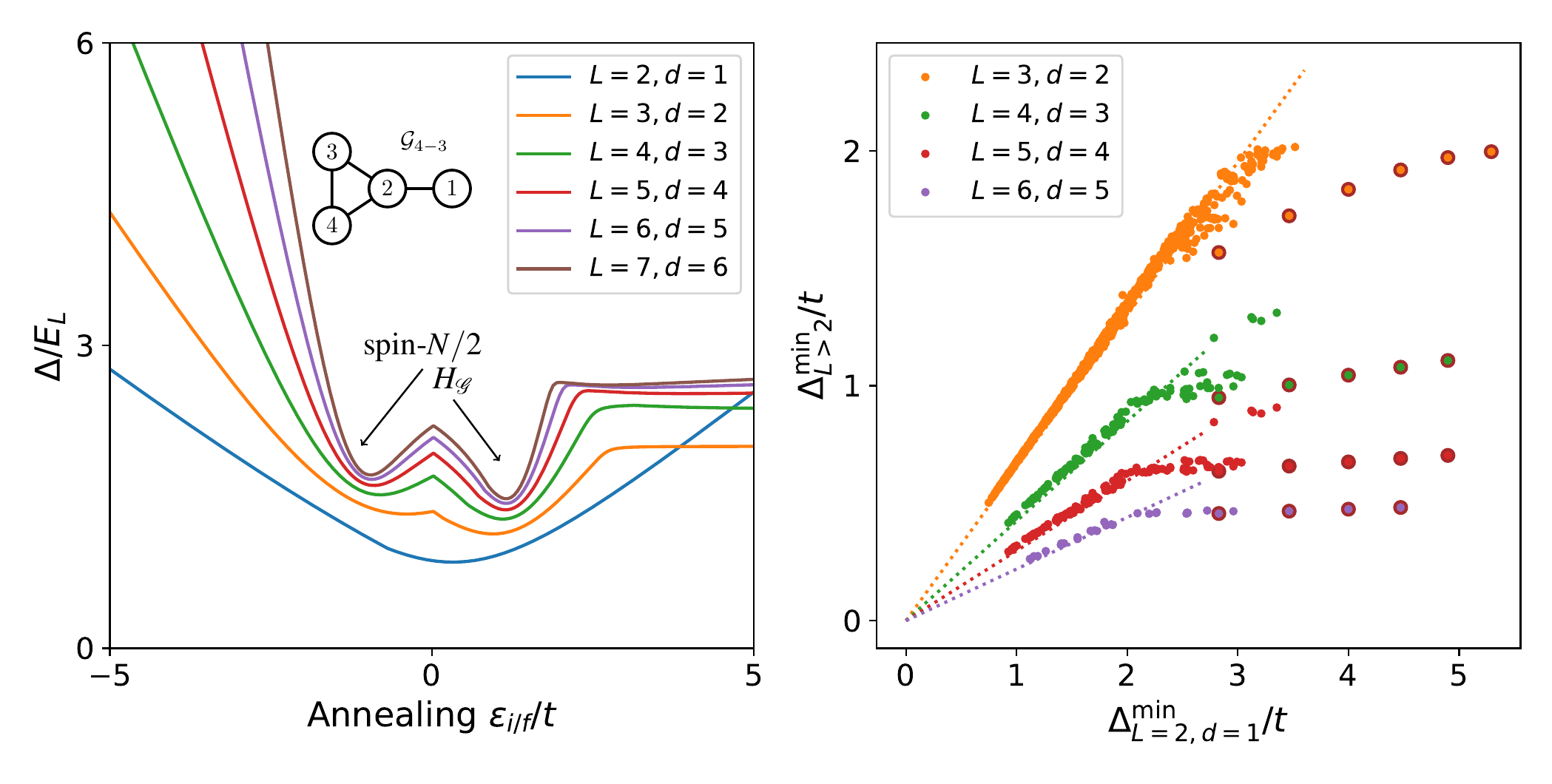}
	\caption{\textbf{Performance of quantum annealing in the hypercube limit $d=L-1$.}
		(a) The gap along the annealing for an implementation encoding the MIS problem on the paw graph $\mathcal{G}_{4-3}$. The two local minima for $L\geq3$ are indicated by arrows. 
		(b) The relation between the minimum gaps for the hypercube limit $\Delta^\mathrm{min}_{L,L-1}$ and the PXP implementation $\Delta^\mathrm{min}_{2,1}$ for all small graphs with $N\le 7$~\cite{read05atlas}. Results are shown for $L=3,4,5,6$ (for graphs with $N=2,3,4,5$), $L=3,4,5$ ($N=6$), or $L=3$ ($N=7$). 
		The dashed lines with slopes $2\sqrt{2}E_{L-1}/\pi t$ are guides to the eyes highlighting the linearity.
	}
	\label{fig:hypercube_limit}
\end{figure}

We proceed to investigate the quantum dynamics on the blockaded chain model along the logical protocol. The hypercube limit, $d=L-1$, exhibits a high degree of connectivity within the Hilbert space. This typically leads to highly delocalized quantum many-body states and improved annealing performance~\cite{cain2023quantum}. Counterintuitively, the annealing performance in this limit is fundamentally determined by the quantum dynamics of the PXP implementation, $L=2,d=1$. Connectivity in Hilbert space configurations alone is therefore insufficient to predict annealing performance.

Along the annealing, the gap exhibits universal behaviors, as depicted for the representative cases in Fig.~\ref{fig:hypercube_limit}. It has one local minimum in the first and second half of the annealing, respectively. Given a specific graph and varying $L\ge 3$, the gaps as a function of the annealing parameters collapse upon a scaling of $L^2\Delta^{\mathrm{min}}_{L,L-1,\mathcal{G}}$. This is shown for the \emph{paw graph}, $\mathcal{G}_{4-3}$, in Fig.~\ref{fig:hypercube_limit}(a).

The universal behaviors can be elucidated through a mapping of $H_{L,L-1,\mathcal{G}}$ onto spinor models, see Appendix~\ref{sec:hypercube_two_mode} for details. The low energy dynamics can effectively be described by three modes along each chain, including two exponential modes, $|A\rangle_n \propto \sum_{\ell=1}^{L} e^{-\epsilon_i \ell/tL} |n,\ell\rangle$, $|C\rangle_n \propto \sum_{\ell=1}^{L} e^{-\epsilon_f (L-\ell)/tL} |n,\ell\rangle$, and a standing-wave mode, $|B\rangle_n \propto \sum_{\ell=1}^{L} \sin(\pi \ell/L) |n,\ell\rangle$.
The first minimum lies around $\epsilon_i=t$, where the system transitions from the localized state $\bigotimes_{n=1}^{N}|A\rangle_n$ to the delocalized state $\bigotimes_{n=1}^{N}|B\rangle_n$. Since any basis state formed by direct products of $|A\rangle_n$ and $|B\rangle_n$ is legitimate, the full Hamiltonian projecting on this basis thus induces the Hamiltonian of a spin-$N/2$, $H^N_{\tilde{L}=2}$, with an effective hopping $\tilde{t} = 4\sqrt{2}/\pi\times \left(1-\cos[\pi/(L+1)]\right) \sim 1/L^2$. 
Similarly for the second local minimum around $\epsilon_{f}=t$, projecting $H_{L,L-1,\mathcal{G}}$ onto the basis states of $|B\rangle_n$ and $|C\rangle_n$ leads to the PXP implementation of the same graph, $H_{\tilde{L}=2,\tilde{d}=1,\mathcal{G}}$. The gap behaviors thus inherit from $H_{2,1,\mathcal{G}}$.

Our mapping can further be confirmed by comparing the minimum gap along protocol between the hypercube limit $d=L-1$ and the PXP implementation $L=2,d=1$, see Fig.~\ref{fig:hypercube_limit}(b). Consistent with our analysis, a strong correlation can be seen between $\Delta^\mathrm{min}_{L,L-1,\mathcal{G}}$ and $\Delta^\mathrm{min}_{2,1,\mathcal{G}}$. During the annealing, the bottleneck of the quantum annealing performance competes between the quantum dynamics of a spin-$N/2$ system, and that of the PXP implementation based on the graph, with $\Delta^\mathrm{min}_{\mathrm{spin}-N/2}=2t$,
\begin{eqnarray}
	\label{eq:hypercube_scaling}
	\Delta^{\mathrm{min}}_{L,L-1,\mathcal{G}} \approx \tilde{t}/t\times \min\left(2t,\,\Delta^\mathrm{min}_{2,1,\mathcal{G}}\right).
\end{eqnarray}
The quantum annealing performance in the hypercube limit is thus fundamentally determined by the quantum dynamics of the PXP implementation. Increasing Hilbert-space connectivity in the hypercube limit does not remove the graph-dependent PXP bottleneck.

\section{Factorizability of quantum states in indent limit}
\label{sec:product_state}

\begin{figure}[!t]
	\centering
	\includegraphics[width=\columnwidth]{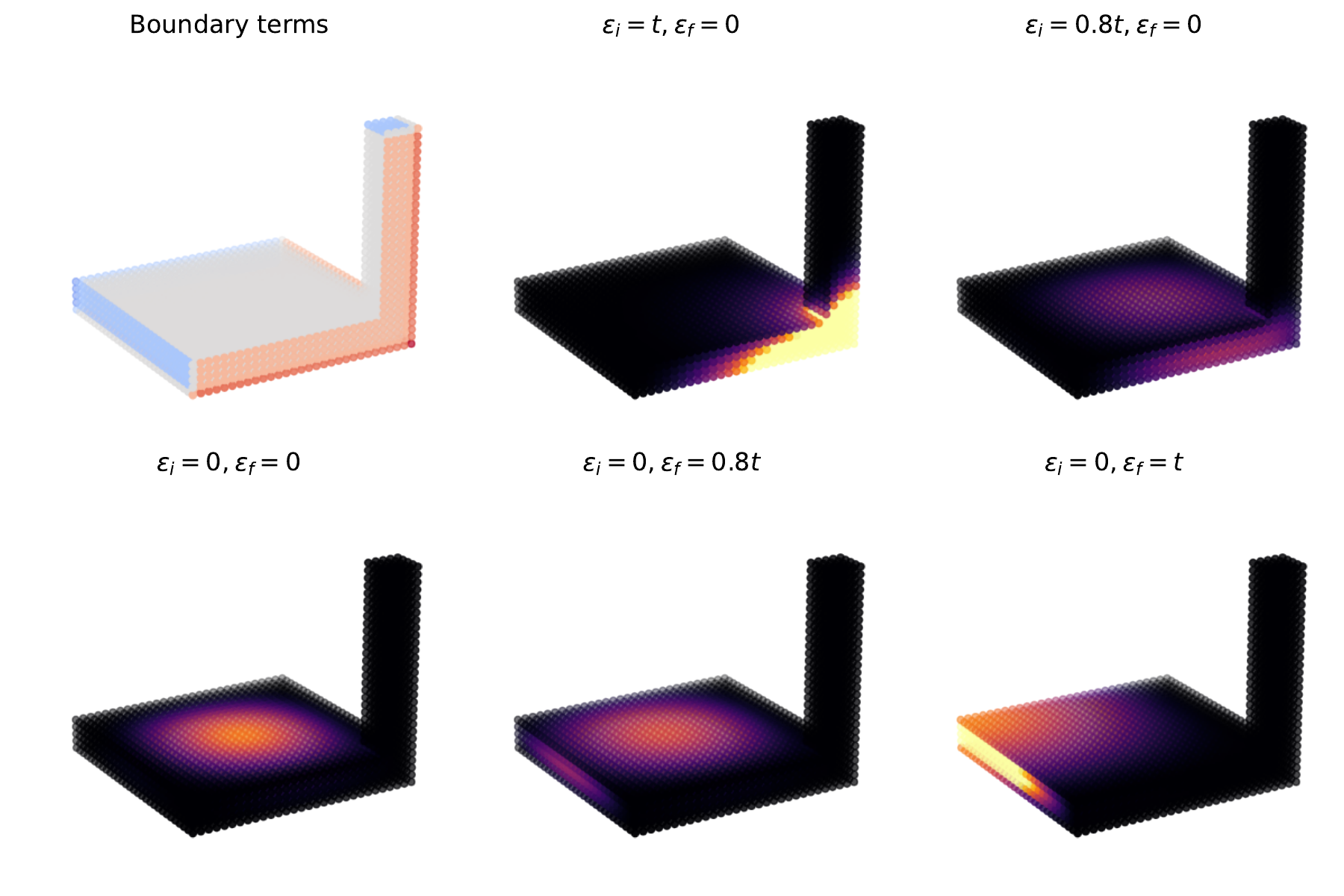}
	\caption{\textbf{Densities of instantaneous ground states at several characteristic time points of the annealing protocol.}
	(a) The lattice sites and boundary terms $\epsilon_i$ (red) and $\epsilon_f$ (blue) for the $(30,5,\mathcal{P}_3)$-blockaded chain model. (b-f) Densities of state along the quantum annealing. The majority of the density lies in the ``flat'' oblate hypercuboid subspace, and abstained from the ``chimney'' prolate subspace. Therefore, the wavefunction can be formulated as a product state.
	}
	\label{fig:slab_density}
\end{figure}

In stark contrast to the hypercube limit, the indent limit of the blockaded-chain model manifests quantum dynamics distinct from the PXP implementation. Importantly, due to the characteristic structure of the Hilbert space, the quantum dynamics becomes tractable and can be attributed to a product-state ansatz. In the rest of this work, we focus on this implementation, and explore the potential for improvement of quantum annealing performance.

In indent limit, the structure of the Hilbert space as described by Eq.~\eqref{eq:hilbert_space_structure} chiefly governs the quantum many-body dynamics. These hypercuboid subspaces can be viewed individually as  quasi-integrable subsystems, which are barely coupled with each others, and host independent dynamics. Specifically, in the low energy sector, the quantum many-body states are thus expected to follow the tentative $N$-partite ansatz based on a mIS (not necessarily optimal), $\sigma^\mathrm{mIS}$, of the graph,
\begin{eqnarray}
	\label{eq:product_state_tent}
	|\Psi^\mathrm{mIS}\rangle_\mathrm{tentative} = \bigotimes_{n=1}^{N} |\psi^{\sigma^\mathrm{mIS}_n}_{\mu}\rangle_n.
\end{eqnarray}
Here, $|\psi^{0}_{\mu}\rangle_n$ ($|\psi^{1}_{\mu}\rangle_n$) are logical zero (one) modes, and corresponds to any vertex with $\sigma^\mathrm{mIS}_n=0$ ($\sigma^\mathrm{mIS}_n=1$) in the mIS. They span a one-dimensional lattice of $d$ ($L$) sites along the corresponding dimension $n$, and are the $\mu$th harmonics of the Hamiltonian $h =  -t \sum_{\ell=1}^{d-1} \left(|\ell\rangle \langle \ell+1| + \mathrm{h.c.}\right) - \epsilon_{i} |1 \rangle\langle 1|$ ($h =  -t \sum_{\ell=1}^{L-1} \left(|\ell\rangle \langle \ell+1| + \mathrm{h.c.}\right) - \epsilon_{f} |L \rangle\langle L| $).

Surprisingly, this convenient fully factorizable ansatz is already sufficient for describing the low-energy quantum dynamics in models based on a large number of graphs. For instance, it is consistent with the instantaneous ground states of $H_{(30,5,\mathcal{P}_3)}$, as shown in Fig.~\ref{fig:slab_density}. However, models based on graphs with permutation symmetries tend to host correlated quantum many-body dynamics. In the following, we progressively demonstrate these correlated effects with exact numerical solutions, and demonstrate examples which are fully factorizable, factorizable with quantum many-body bound states, and nontrivially correlated.

\subsection{Completely factorizable quantum states}

\begin{figure}[!t]
	\centering
	\includegraphics[width=\columnwidth]{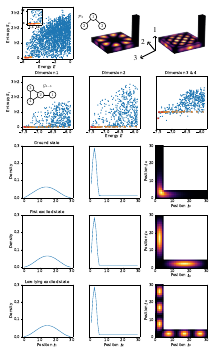}
	\caption{{\bf Product-state ansatz with homogeneous boundary.} (a) The entanglement spectrum of the system at midpoint of protocol, where the boundary is homogeneous $\epsilon_{i,f}=0$, for graph $\mathcal{P}_3=\mathcal{G}_{3-1}$. The entanglement entropy $S_1$, cf. Eq.~\eqref{eq:def_entropy}, is evaluated for each eigenstate. The low energy sector is dominated by a large family of product states with vanishing entanglement entropy, as marked by the orange crosses. (b,c) The densities of two arbitrarily chosen zero-entanglement states, marked by red crosses in panel (a).  (d-f) The entanglement spectrum for the paw graph, $\mathcal{G}_{4-3}$. (g-o) The projection of (g-i) the ground state, (j-l) the first excited state, and (m-o) a low-lying excited state is consistent with the product state ansatz. Along the \nth{3} and \nth{4} dimensions, the state is either (l,o) a superposition of two harmonics or (i) a bound state.
	}
	\label{fig:product_ansatz}
\end{figure}

We first examine the validity of the tentative ansatz at the homogeneous boundary condition $\epsilon_{i,f}=0$. We mainly focus on the separability of the state and its projected densities along different dimensions. The separability of the quantum states can be quantified by its entanglement entropy. Given many-body state $|\Psi\rangle$ based on a lattice of $N$ dimensions and length $L$, its entanglement entropy with respect to a specific dimension $n$ can be defined as
\begin{eqnarray}
	\label{eq:def_entropy}
	S_n &=& -\sum_{\ell=1}^{L} \lambda_\ell \ln \lambda_\ell, \quad  \lambda_\ell \in \mathrm{eig}(\rho_n)
\end{eqnarray}
where the reduced density matrix
\begin{eqnarray}
	\label{eq:def_rho_n}
	\rho_n = \tr_{m\neq n}|\Psi\rangle\langle\Psi|
\end{eqnarray}
is traced over all $N-1$ dimensions other than $n$.

For the case of 3-chain graph $\mathcal{P}_3=\mathcal{G}_{3-1}$, and an implementation with $L=30,d=5$, we show the entanglement spectrum with respect to the \nth{1} dimension in Fig.~\ref{fig:product_ansatz}(a). The low-energy sector is indeed dominated by a family of states with zero entanglement, confirming the states are product states. Two arbitrary zero-entanglement excited states are visualized in Fig.~\ref{fig:product_ansatz}(b,c), where their densities apparently exhibit higher harmonics among different dimensions. These states are also characterized by $S_{2/3}=\ln 2$ due to (anti)symmetric superposition.

Intuitively, similar arguments apply straightforwardly to the case of paw graph $\mathcal{G}_{4-3}$, apart from its ground state. In Fig.~\ref{fig:product_ansatz}(d-f), we show the entanglement spectra along different dimensions. Similar to the $\mathcal{P}_3$, there is a large family of states with $S_1=S_2=0$, and $S_3=S_4=\ln2$, which confirms that they are product states. This can be further confirmed by the projected densities along different dimensions, as shown in Fig.~\ref{fig:product_ansatz}(g-o). Along the \nth{1} and \nth{2} dimensions, the projected densities correspond to the basis states $|\psi^{1}_{1}\rangle_n$ and $|\psi^{0}_{1}\rangle_n$, respectively. Particularly, $|\psi^{1}_{\mu}\rangle_n = \sum_{\ell=1}^{L} \sin(\mu\pi \ell/L) |n,\ell\rangle$ and $|\psi^{0}_{\mu}\rangle_n = \sum_{\ell=1}^{d} \sin(\mu \pi \ell/d) |n,\ell\rangle$. Along the \nth{3} and \nth{4} dimensions, the projected densities indicate that the first excited state [Fig.~\ref{fig:product_ansatz}(l)] and the low-lying excited state [Fig.~\ref{fig:product_ansatz}(o)] are superposition states $(|\psi^{1}_1\rangle_3|\psi^{1}_3\rangle_4\pm|\psi^{1}_3\rangle_3|\psi^{1}_1\rangle_4)/\sqrt{2}$.

The ground state, however, can not be fully captured by the tentative ansatz. This is the most obvious in its projected density along the \nth{3} and \nth{4} dimensions. It resembles the bound state which appears in the model based on the 2-vertex graph $\mathcal{P}_2=\mathcal{G}_{2-1}$, as studied in Ref.~\cite{bombieri25quantum}. Correspondingly, its entanglement entropy along these two dimensions is $0<S_{3/4}<\ln 2$ [Fig.~\ref{fig:product_ansatz}(f)]. This example highlights the broad existence and importance of bound state dynamics in blockaded-chain models.

\subsection{Clique-based bound states as factor basis states}

\begin{figure}[!t]
	\centering
	\includegraphics[width=\columnwidth]{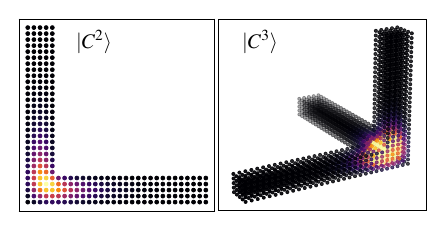} 
	\caption{{\bf Quantum bound states associated to cliques.} Densities of clique states $|K^q\rangle$ with homogeneous boundary $\epsilon_{i,f}=0$ with $L=30,d=5$ for $q=2,3$.
	}
	\label{fig:clique}
\end{figure}

\begin{figure}[!t]
	\centering
	\includegraphics[width=\columnwidth]{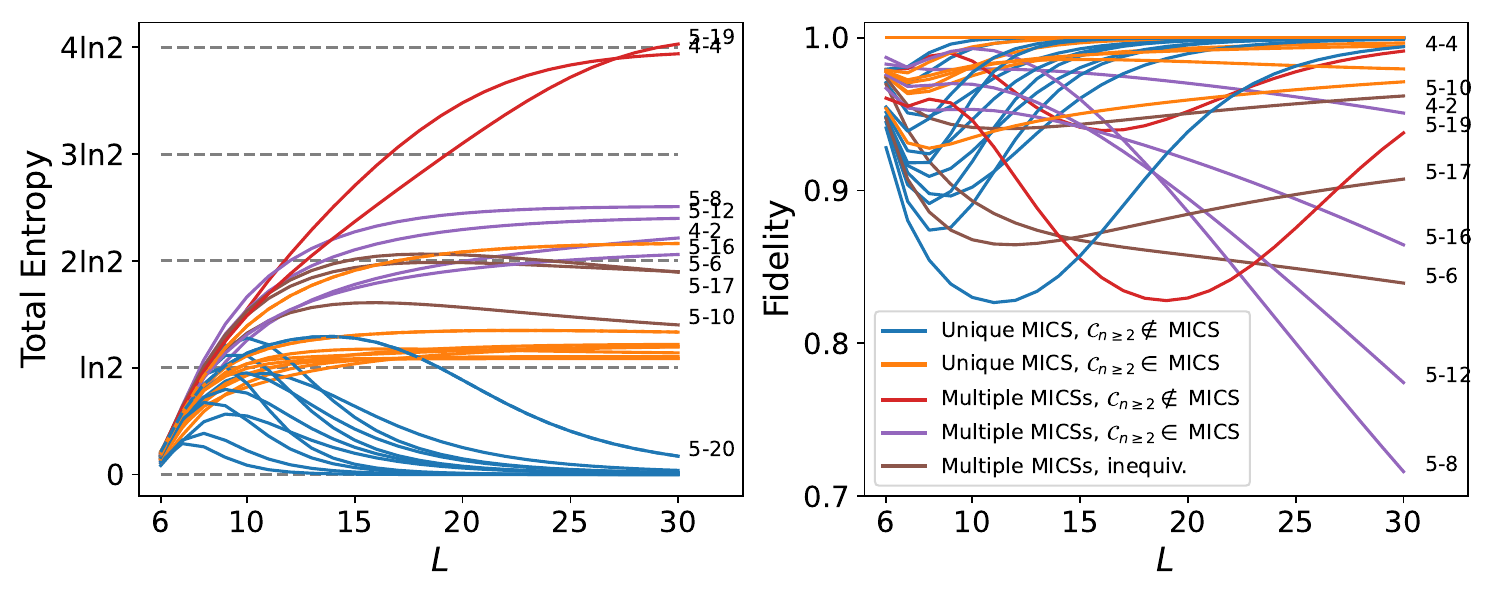}
	\caption{{\bf Product-state ansatz for small graphs.}
		(a) Total entanglement entropy and (b) fidelity with respect to the MICS-based product-state ansatz, Eq.~\eqref{eq:product_state}, of the ground state with homogeneous boundary, $\epsilon_{i,f}=0$, for different $L$ at fixed $d=5$. A collapse in the behaviors can be observed in both observables. The curves are categorized into five groups according to the properties of the MICSs of the graphs for a clear presentation.
	}
	\label{fig:entropy_fidelity}
\end{figure}

The bound states are typically associated to complete subgraphs, or cliques. This motivates us to investigate the blockaded chain model based on complete graphs $\mathcal{K}_{q\ge2}$, e.g.\ $\mathcal{G}_{2-1}$, $\mathcal{G}_{3-2}$, $\mathcal{G}_{4-6}$, $\mathcal{G}_{5-21}$. Their Hilbert spaces have the shapes of a ``claw'', which is composed by a ``palm'', $[1,d]^q$, and $q$ ``fingers'', $[1,d]^{q-1}\times [d+1,L]$. The logical-zero sector of volume $d^q$ acts like a base, and $q$ logical-one sectors of volume $L\times d^{q-1}$ are hypertubes individually attached to the based. Inherited from the graph, the Hilbert space also has a permutation symmetry of degree $q$. Specifically, the two-dimensional case of a $\mathsf{L}$-shaped lattice has been studied thoroughly in Ref.~\cite{bombieri25quantum}, where a bound state of particularly low energy has been seen and attributed to quantum interference.

With homogeneous boundary, the existence of similar bound states in higher dimensions can be argued using energetic arguments for the ground states $|K^q\rangle$ of $H_{L,d,\mathcal{K}_{q}}$. The variational principle for ground states dictates a strictly monotonic decrease in their energies, $E_q$,
\begin{eqnarray}
	\label{eq:inequality_variation}
	E_q - qE_0 < E_{q-1} - (q-1)E_0 < \dots < E_2 - 2E_0, \nonumber \\
\end{eqnarray}
where $E_0/t=-2\cos[\pi/(d+1)]$ is the energy of the first harmonic mode spanning $d$ sites. This verifies that the ground states of $H_{L,d,\mathcal{K}_{q}}$ are indeed exponentially decaying along the ``fingers'', since $E_q - (q-1) E_0 < E_2 - E_0< -2t$~\cite{bombieri25quantum}, and localized around the ``palm'', i.e., the logical-zero sector. The two-dimensional and three-dimensional bound states, $|K^2\rangle$ and $|K^3\rangle$, are visualized in Fig.~\ref{fig:clique}. 

The ansatz for the quantum many-body state should thus take these bound states as indivisible factors. For instance, with homogeneous boundary, $\epsilon_{i,f}=0$, we propose the following improved ansatz represented by an independent clique set (ICS), $\mathcal{K}=\{K_r\}$, and verify its fidelity against the exact quantum many-body ground state,
\begin{eqnarray}
	|\Psi^\mathrm{ICS}\rangle = \bigotimes_{r=1}^{R} |K^{q_r}\rangle_{n\in K_r} \otimes \bigotimes_{n\notin \mathcal{K}} |\psi^0\rangle_n.
\end{eqnarray}
Here, $|\psi^0\rangle_n = \sum_{\ell=1}^{d} \sin(\pi \ell/d) |n,\ell\rangle$, $|K^1\rangle_n = \sum_{\ell=1}^{L} \sin(\pi \ell/L) |n,\ell\rangle$, and $|K^{q\ge2}\rangle_{n_1,\dots n_q}$ are the $q$-dimensional bound states associated to $K_M$.

We examine the ansatz for all small graphs with $N\le5$ vertices in Fig.~\ref{fig:entropy_fidelity} by computing their sum of single-dimension entropies $S=\sum_{n=1}^{N} S_n$ and fidelity with respect to the ansatz $F=|\langle \Psi^\mathrm{GS}| \Psi^\mathrm{MICS}\rangle|$, for varying $L$ at fixed $d=5$. Particularly, when the graph possesses multiple MICSs, the equal-weight superposition is considered, $|\Psi^\mathrm{MICS}\rangle \propto |\Psi^\mathrm{MICS1}\rangle+|\Psi^\mathrm{MICS2}\rangle+\cdots$. 

The behaviors of the quantum state is fully consistent with the ansatz when the graph has a unique MICS (blue, orange), or when the graph has multiple MICSs but all the MICSs are composed by single vertices as cliques (red). See Table~\ref{tab:graph_list} in Appendix~\ref{sec:graph_list} for a list of graphs.
In these cases, the fidelity converges to $1$ for large $L$, while the entropy converges to values reflecting the bound states and/or the superposition states. For example, for large $L$, $S_{5-1}=0$ because $\mathcal{G}_{5-1}$ has a MICS of four individual vertices, $S_{5-2}=2S_{\mathcal{C}_2}$ because $\mathcal{G}_{5-2}$ has a MICS consisting of a 2-clique, and $S_{4-4}=4\ln2$ because $\mathcal{G}_{4-4}$ has a MICS consisting two pairs of equivalent vertices. 

However, the quantum states go beyond the MICS product-state ansatz when the graph possesses multiple MICSs consisting of cliques of $M\ge2$ vertices. These MICSs can be either  equivalent, e.g., $\mathcal{G}_{4-2}$, $\mathcal{G}_{5-8}$, $\mathcal{G}_{5-12}$, and $\mathcal{G}_{5-16}$, or inequivalent to each other, e.g., $\mathcal{G}_{5-6}$, $\mathcal{G}_{5-10}$, and $\mathcal{G}_{5-17}$. These graphs are characterized by high permutation symmetry, cf.\ Table~\ref{tab:graph_list} in Appendix~\ref{sec:graph_list}. In the quantum many-body system, this symmetry is inherited in the Hilbert space structure, and induces hybridization and binding effects. This leads to strong quantum correlations beyond the ansatz.

\section{Hybridization and binding}
\label{sec:analysis}

The numerical evidences indicate that the ground state of the system at homogeneous implementation deviates from the tentative product state ansatz, Eq.~\eqref{eq:product_state_tent}, when the symmetry of the graph is high. We proceed to formally investigate how the interplay between symmetry of the Hilbert space and geometric constraints leads to hybridization and binding of the product states, and induces quantum  correlations and potentially bound states.

\subsection{Binding in the $\mathcal{K}_q$  graphs}
\label{sec:binding_scattering}

The bound states in complete graphs, $\mathcal{K}_q$, constitute the first effects of symmetry. They stem from coordinated enhancement of equivalent scattering channels induced by symmetry. The energy of the logical-zero state, $|\psi^0\rangle^{\otimes q}$, is significantly lowered and becomes the ground state of the system.

The Hamiltonian $H_{\mathcal{K}_q}$ couples $|\psi^0\rangle^{\otimes q}$ in the ``palm'' to lattice sites in the ``fingers'', $H |\psi^0\rangle^{\otimes q} = q E_0 |\psi^0\rangle^{\otimes q} - \tilde{t} \sqrt{q} |B\rangle$, where $|B\rangle = \frac{1}{\sqrt{q}} \sum_{n=1}^{q} |n,d+1\rangle\otimes |\psi^0\rangle^{\otimes (q-1)}$ is the ``border state'', and where $\tilde{t} = t \psi^0(d)$ is the hopping amplitude on the boundary, with $\psi^0(d)=\sqrt{2/(d+1)}\sin[\pi d/(d+1)]$ the wavefunction on site $d$. The border state $|B\rangle$ thus provides a scattering channel composed symmetrically from $q$ equivalent and independent channels along the ``fingers''.

The scattering channels effectively induce a self energy upon the bare $|\psi^0\rangle^{\otimes q}$ state,
\begin{eqnarray}
	\Sigma = (q-1)\tilde{t}^2 \langle d+1 | G | d+1 \rangle.
\end{eqnarray}
Here, $G = 1/[E-(q-1)E_0-h]$ is the propagator along a one-dimensional lattice with Hamiltonian $h$ and energy shifted by $(q-1)E_0$ due to contribution from the $(q-1)$ $|\psi^0\rangle$ factors in $|B\rangle$. The propagator further has an overall factor of $(q-1)$ instead of $q$, which appears as an effective reduction of scattering channels. 

The energy of the dressed $|\psi^0\rangle^{\otimes q}$ state can thus be determined by  the self-energy equation as $E \approx (q-1) E_0 - 2t \cosh(\pi\alpha_q/(d+1))$, with
\begin{eqnarray}
	\label{eq:bound_energy_alpha}
	1+\alpha_q^2= \frac{q-1}{\pi \alpha_q} \left(1-e^{-2\pi\alpha_q}\right).
\end{eqnarray}
Its energy is reduced to below the threshold $E<(q-1)E_0-2t$, and it is thus a bound state with wavefunctions decaying along the one-dimensional lattice.

Bound states are thus correlated quantum many-body states induced by symmetry and geometric constraints. Symmetry induces dressing through scattering channels to the bare logical-zero state, $|\psi^0\rangle^{\otimes q}$ and significantly lowers its energy, while geometric constraints increase the energy cost for a propagating state. Detailed derivations are available in Appendix~\ref{sec:deformation_MICS_appendix}.

\subsection{Hybridization in the $\mathcal{P}_4$ graph}

Besides coordinated enhancement, symmetry leads to deviation to the product state ansatz through hybridization. The $N=4$ path graph, $\mathcal{P}_4=\mathcal{G}_{4-2}$, has two equivalent MICSs, each of them consisting of a vertex and a 2-clique. It provides an example how coherent hybridization between the two symmetry-related MICS product states deforms the constituent pair wavefunctions.

The product state ansatz predicts two basis states related by the reflection symmetry $(1,2,3,4)\leftrightarrow (4,3,2,1)$, formally written as $|\Psi_\uparrow\rangle = |A\rangle_{1,2}\otimes|B\rangle_{4,3}$ and $|\Psi_\downarrow\rangle = |B\rangle_{1,2}\otimes|A\rangle_{4,3}$. Specifically for $\mathcal{P}_4$, these components are $|A\rangle_{m,n} = |\psi^1\rangle_m \otimes|\psi^0\rangle_n$ and $|B\rangle_{m,n}=|K^2\rangle_{m,n}$. The basis state $|\Psi_{\uparrow}\rangle$ ($|\Psi_{\downarrow}\rangle$) is an exact eigenstate of the Hamiltonian $P_2 H_{\mathcal{P}_4} P_2$ ($P_3 H_{\mathcal{P}_4} P_3$), where $P_n$ projects out the non-MICS lattice sites $|n,\ell\ge d+1\rangle$. Nevertheless, applying the bare Hamiltonian on these states leads to leakage into the non-MICS Hilbert subspace, $\vert \langle 2/3, \ell=d+1| H_{\mathcal{P}_4}|\Psi_{\uparrow/\downarrow}\rangle \vert > 0$.

To appreciate the role of the reflection symmetry, we additionally consider the paw graph $\mathcal{G}_{4-3}$, which has an extra edge $<2,4>$ on top of $\mathcal{P}_4$. This graph has a single MICS, and $|\Psi_\uparrow\rangle$ is the product state ansatz for its unique ground state. This leakage $H_{\mathcal{G}_{4-3}}|\Psi_\uparrow\rangle$ is negligible compared to the energy difference between the MICS state and the suboptimal ICS state, $|\psi^0\rangle_1|\psi^1\rangle_2|\psi^0\rangle_3|\psi^0\rangle_4$. According to perturbation theory, $|\Psi_\uparrow\rangle$ thus serves a good approximation for the ground state in the indent and continuum limit, $L\gg d\gg 1$.

For $\mathcal{P}_4$, however, the reflection symmetry leads to degenerate basis product states. The tiny leakage can potentially induce significant deformation to the ground state while lifting the degeneracy. As a result, the exact ground state behaves differently from the symmetric superposition $|\Psi_+\rangle \propto |\Psi_\uparrow\rangle+|\Psi_\downarrow\rangle$. We capture the effects of reflection symmetry by expanding the quantum state in the basis,
\begin{eqnarray}
	\label{eq:basis_states_P4}
	\{\Pi |\Psi_{++}\rangle, \, \Pi (|\Psi_{+-}\rangle + |\Psi_{-+}\rangle)/\sqrt{2}, \, \Pi |\Psi_{--}\rangle\}.
\end{eqnarray}
Here, $\Pi = P_{<2,3>}$ [cf.\ Eq.~\eqref{eq:projection_operator}] describes the geometric constraint between the \nth{2} and \nth{3} dimensions. $|\Psi_{\rho\sigma}\rangle = |u_\rho\rangle_{1,2}\otimes |u_\sigma\rangle_{4,3}$, with $\rho=\pm$, $\sigma=\pm$. And $|u_\pm\rangle_{m,n} = \frac{|A\rangle_{m,n} \pm |B\rangle_{m,n}}{\sqrt{2(1\pm s)}}$ with $s=\langle A|B\rangle$. This basis is chosen, because the non-projected wavefunctions $|\Psi_{\mu\nu}\rangle$ are orthonormal and nearly resonant with each other. The antisymmetric superposition $|\Psi_{+-}\rangle - |\Psi_{-+}\rangle$ does not couple into the basis. Particularly, the symmetric product state can be written as,
$|\Psi_+\rangle = \frac{1+s}{\sqrt{2(1+s^2)}}|\Psi_{++}\rangle - \frac{1-s}{\sqrt{2(1+s^2)}}|\Psi_{--}\rangle$.

\begin{figure}[!t]
	\centering
	\includegraphics[width=\columnwidth]{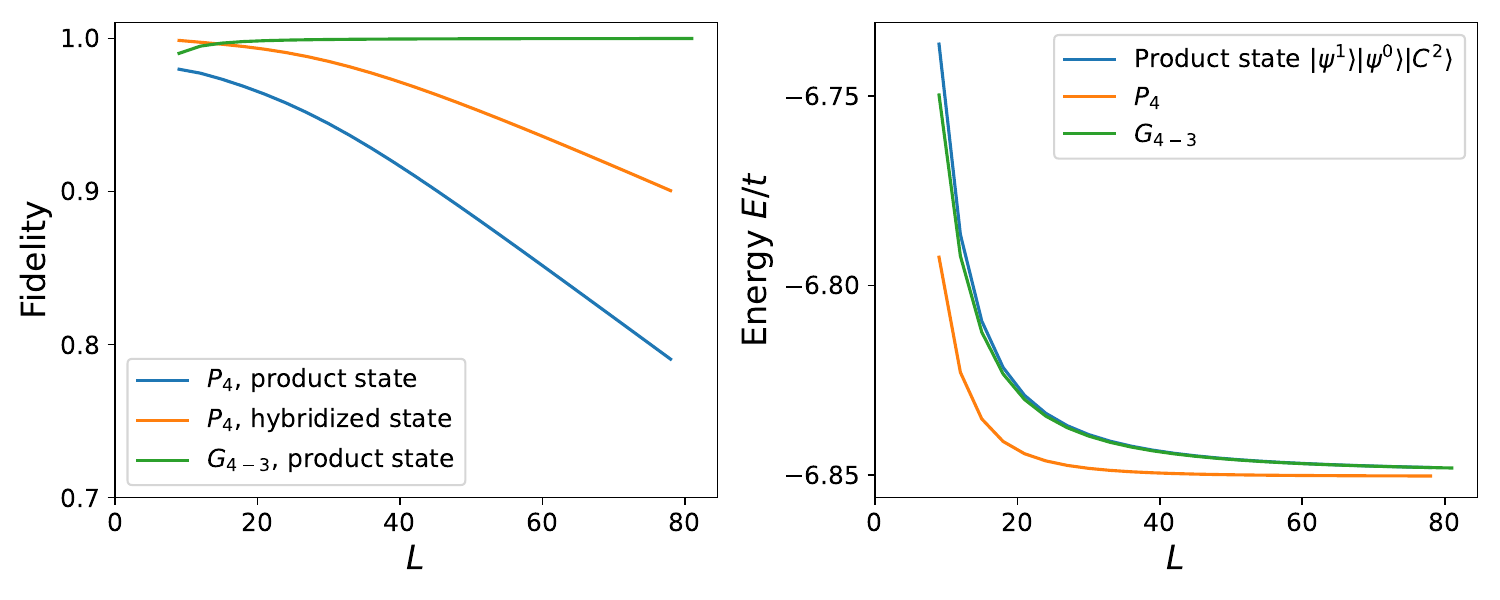} 
	\caption{{\bf Hybridization of quantum state in $\mathcal{P}_4$ graph.} (a) The fidelity of the product state ansatz, Eq.~\eqref{eq:product_state}, and the hybridized ansatz, Eq.~\eqref{eq:G43_state_solution} with respect to the exact quantum state for $\mathcal{P}_4$ graph at homogeneous boundary. (b) The energy of the exact $\mathcal{P}_4$ state compared to the energy of the product state ansatz. The gap between them vanishes in indent limit. For comparison, the control case of paw graph $\mathcal{G}_{4-3}$ is also shown, where the product state ansatz is precise up to negligible perturbation.
	}
	\label{fig:hybrid}
\end{figure}

The Schr\"odinger equation projected onto this basis, Eq.~\eqref{eq:basis_states_P4}, is  $H_{\Pi} \psi = E S \psi$, with $(H_\Pi)_{\mu\nu} = \langle \Psi_\mu | H | \Psi_\nu\rangle$ and $S_{\mu\nu} =  \langle \Psi_\mu | \Psi_\nu\rangle$ the Gram matrix. The indices $\mu,\nu$ run over the three basis states, and the ground state $\psi=(\psi_{++},\psi_{+-},\psi_{--})^T$ is normalized as $\psi^\dagger S \psi = 1$. 
The system thus hosts deformed eigenstate,
\begin{subequations}
	\label{eq:G43_state_solution}
	\begin{eqnarray}
		|\Psi\rangle &=& \frac{1}{2(1-s^2)}\Pi |\Psi'\rangle\\
		|\Psi'\rangle &=& \left(\beta+\gamma\right) |A\rangle_{1,2}\otimes|A\rangle_{4,3}   + \left(\beta-\gamma\right) |B\rangle_{1,2}\otimes|B\rangle_{4,3} \nonumber \\
		&& - \delta \left(|A\rangle_{1,2}\otimes|B\rangle_{4,3}+|B\rangle_{1,2}\otimes|A\rangle_{4,3}\right)
	\end{eqnarray}
\end{subequations}
where $\beta=(\psi_{++}+\psi_{--}) - s(\psi_{++}-\psi_{--})$, $\gamma=\psi_{+-}\sqrt{2(1-s^2)}$, $\delta= s(\psi_{++}+\psi_{--}) - (\psi_{++}-\psi_{--})$.

In the absence of the projection operator $\Pi$, the three basis states in Eq.~\eqref{eq:basis_states_P4} remain orthonormal and decoupled by construction. The geometric constraint imposed by the projector $\Pi$ thus introduces overlap of the basis states through $S\neq \mathds{1}$, as well as a coupling between them through $H_{\Pi}\neq 0$. This leads to non-perturbative hybridization of the quantum state, and thus deviation from the product state ansatz.

We quantify this geometric deformation by considering $d=3$ and $L$ varying from $10$ to $80$, see Fig.~\ref{fig:hybrid}(a). We evaluate the fidelity of the exactly solved quantum state against the symmetric composition of the product state $|\Psi_+\rangle$, and the state solved within the three-state basis. Both fidelities deteriorate as $L$ increases, signaling an enhanced hybridization. This indicates that the hybridization is non-perturbative in indent limit. However, the energy of the exact quantum state converges to the product state ansatz in indent limit, see Fig.~\ref{fig:hybrid}(b). The decreasing gap indicates that, even though hybridization brings the energy of the state to below the propagating threshold, it does not lead to a bound state, since the nominal decaying length extends with the system length. More importantly, unlike binding, hybridization has no energetic contribution in indent limit. As a control comparison, the fidelity and energy for the $\mathcal{G}_{4-3}$ quantum state are also shown. Indeed, due to the lack of degeneracy, the deviation of the exact quantum state from the product state ansatz is perturbative.

Apart from complete graphs, genuine bound states are also found in the 5-cyclic graph $\mathcal{G}_{5-8}=\mathcal{C}_5$, and the house graph $\mathcal{G}_{5-12}$ among all graphs with $N=5$ vertices, cf.\ Table~\ref{tab:graph_list} in Appendix~\ref{sec:graph_list}. These graphs are later referred to \emph{binding motifs}, and are all collected in the set $\mathcal{G}_\mathrm{bound}=\{\mathcal{P}_2,\mathcal{K}_3,\mathcal{K}_4,\mathcal{C}_5,\mathcal{G}_{5-12},\mathcal{K}_6,\dots\}$. Cliques constitute a major family of motifs.

\subsection{Product-state ansatz based on cliques and motifs}
\label{sec:ansatz}

Bound states are generically associated to graphs with high symmetry, particularly complete graphs. They urge a modification to the tentative product-state ansatz for a comprehensive description of quantum dynamics along annealing in indent limit. Particularly, since hybridization induces weak energetic effects in indent limit, it is not explicitly taken into account in the ansatz.

The quantum many-body state can be represented by a generic IS, $\mathcal{S} = \{ n \mid  \sigma^\mathrm{IS}_n = 1\} \subset \mathcal{V}$, which is not necessarily maximal. Along the dimensions $n\in \mathcal{S}$ corresponding to occupied vertices, the state hosts the logical-one mode $|\psi^1\rangle$ generated by a chain of $L$ sites with boundary term $\epsilon_f$. Particularly, we denote the ground state and first excited state for logical-one mode for the instantaneous boundary condition as $|\psi^1_0\rangle$ and $|\psi^1_1\rangle$, respectively. These notations are convenient for discussion on gap behaviors.

The rest of the vertices $\mathcal{V}\setminus\mathcal{S}$ is further grouped into motifs $M_r\in\mathcal{G}_\mathrm{bound}$, $r=1,\dots,R$, such that the union $\mathcal{M} \cup \mathcal{S}$ is an independent motif set with $\mathcal{M} = \{M_r\} \subset \mathcal{V}\setminus\mathcal{S}$. The quantum state ansatz then associates each motif to its corresponding basis state $|M_r\rangle$, subject to the boundary term $\epsilon_{i}$. A motif-based bound state represents a classical configuration where the corresponding vertices are logical-zeros.

The final remaining vertices in $\mathcal{U}=\mathcal{V}\setminus(\mathcal{S}\cup\mathcal{M})$ are then assigned with the logical-zero mode $|\psi^0\rangle$. This is the ground state of a chain of $d$ sites subject to boundary term $\epsilon_{i}$. The factor states $|\psi^0\rangle$ and $|\psi^1\rangle$ are identical to the ones entering the tentative ansatz, Eq.~\eqref{eq:product_state_tent}.

The comprehensive product state ansatz associated to $\mathcal{S}$ and $\mathcal{M}$ is thus
\begin{eqnarray}
	\label{eq:product_state}
	|\Psi\rangle = \bigotimes_{n\in\mathcal{S}} |\psi^{1}\rangle_n \otimes \bigotimes_{n\in \mathcal{U}} |\psi^{0}\rangle_n 
	\otimes  \bigotimes_{r=1}^{R}|M_r\rangle_{n\in M_r}. \nonumber \\
\end{eqnarray}
In the following, unless explicitly specified, we focus on the effects of cliques, $|K^{q\ge2}\rangle$, among all binding motifs. This improves the clarity of our discussion without losing the essential features of our system. The factor states are summarized as follows, 

\begin{eqnarray}
	\label{eq:factor_states}
	\begin{tabular}{c|c|c|c}
		\hline
		Factor & Bulk & Boundary & Logic\\
		\hline\hline
		$|\psi^0\rangle$ & 1D chain of length $d$ & $\epsilon_{i}$ & $0$ \\
		\hline
		$|\psi^1_{0/1}\rangle$ & 1D chain of length $L$ & $\epsilon_{i}$,$\epsilon_{f}$ & $1$ \\
		\hline
		$|M\rangle$ & $|M|$-D lattice from motif $M$ & $\epsilon_{i}$ & $0^{\otimes |M|}$ \\
		\hline
		$|K^{q\ge2}\rangle$ & $q$-D lattice from $\mathcal{K}_q$ & $\epsilon_{i}$ & $0^{\otimes q}$ \\
		\hline
	\end{tabular} 
	\nonumber \\
\end{eqnarray}

\subsection{Energetics for solving MIS}

With homogeneous boundary, the quantum blockaded-chain model thus selects the independent set energetically. To identify the ground state, we associate the cliques with weights determined by the energies of the bound states.  The energy of a product state is $E = \sum_{r=1}^{R} E_{r}+ N_0E_0$, where $E_r$ is the energy of the bound state $|K^{q_r}\rangle$, and $N_0=N-\sum_{r=1}^R q_r$ is the number of vertices not belonging to any clique. 
We can thus assign an effective weight to these cliques to account for this energy, 
\begin{eqnarray}
	\label{eq:weight}
	W_q = \left\vert \frac{E_M-qE_0}{E_1-E_0} \right\vert.
\end{eqnarray}
The weight $W_q$ increase strictly monotonically with $q$, cf.\ Inequality~\eqref{eq:inequality_variation}, and decreases as the shape ratio $d/L$ approaches $0$, i.e., the indent limit. They converge to $W_q^\infty \approx 1 + \alpha_q^2$, $q\ge2$ in indent limit, c.f., Eq.~\eqref{eq:bound_energy_alpha}, with particularly $W^\infty_2\sim1.071$~\cite{lenz86quark}. These weights measure the boundedness of the quantum states, and provide an effective count of vertices inside the clique. At homogeneous boundary fields, maximizing the clique weight minimizes the energy within the clique-product ansatz. The ground state thus search for the ICS with the maximum effective number of vertices, i.e., the MWICS of the graph. A weight can be assigned to a general binding motif by replacing $q$ by $|M|$.

We further investigate the performance of quantum annealing for a specific set of graphs. For clarity of discussion, we consider the family of triangle-free graphs. Per definition, these graphs do not host cliques with $M\ge3$ vertices. The dominant contributions from $\mathcal{G}_\mathrm{bound}$ are thus a pair of connected vertices, $\mathcal{P}_2$. Besides, the graphs can also possibly host a binding motif of $\mathcal{C}_5\in\mathcal{G}_\mathrm{bound}$, or any other unexplored binding motifs with $M\ge6$ vertices. For simplicity of the discussion, we explicitly omit the contributions from these binding motifs momentarily. Effectively, the blockaded-chain model thus searches for the dissociation set $\mathcal{D}$ which maximizes $N_1+W_2^\infty N_2$. In graph theory, a dissociation set contains $N_1$ single vertices and $N_2$ vertex pairs, which are independent from each other.

We clarify whether and when the effective optimization problem facilitates the search for MIS in triangle-free, cubic graphs. These graphs have lower bound for its independence number, $\alpha(G) \ge \frac{3}{10} N$~\cite{dvorak14subcubic}; and elementary edge counting yields an upper bound for its dissociation set, $6N_1+10N_2\le3N$~\cite{diestel25graph}. A failure for the effective optimization problem to reproduce the MIS solution can happen when $N_2\ge \lceil 1/(W_2^\infty-1)\rceil=15$. In other words, the MIS solution is produced for a triangle-free, cubic graph with $N\le64$ vertices; and a suboptimal solution with $\lfloor N/64 \rfloor$ vertices fewer than MIS can potentially be produced for a larger graph.
Moreover, if the graph is further planar, $\alpha(G) \ge \frac{5}{14} N$~\cite{batenburg20large}, this threshold is increased to $N\le 72$. Indeed, the important index is the density of the independent set, $\alpha(G)/N$, since a tradeoff between single vertex and vertex pair becomes hindered by constraint with a dense MIS. If the MIS occupies more than half of the vertices of the graph, $\alpha(G)/N\ge1/2$, the effective optimization produces the correct MIS solution. Our statements particularly apply for cubic graphs with girth at least 6, since cyclic graphs with $N\ge6$ vertices are not associated to bound states, $\mathcal{C}_{N\ge6}\notin \mathcal{G}_\mathrm{bound}$.

\subsection{Dominant product-state channel and clique-controlled gaps}
\label{sec:factorizable}

\begin{figure}[!t]
	\centering
	\includegraphics[width=\columnwidth]{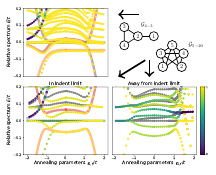}
	\caption{{\bf Representative low-energy spectra in indent limit.} 
	(a) The spectrum of the blockaded-chain Hamiltonian based on the paw graph $\mathcal{G}_{4-3}$ with $L=20,d=3$, compared against the energies of product states estimated by the ansatz Eq.~\eqref{eq:product_state}. Specifically, the orange curve represents the $(1_0,0,0,0)$ state, red $(1_1,0,0,0)$, green $(1_0,0,1_0,0)\pm (1_0,0,0,1_0)$, purple $(0,1_0,0,0)$, and brown $(1_1,0,1_0,0)\pm (1_1,0,0,1_0)$. The color scheme encodes $P_T$, cf.\ Eq.~\eqref{eq:def_PT}, and reflects whether the quantum state lies in the Hilbert subspace of correct MIS.
	(b,c) The spectra based on the $\mathcal{K}_5\setminus e$ graph, $\mathcal{G}_{5-20}$, (b) in indent limit $L=20,d=3$ and (c) away from indent limit $L=20,d=10$. In (b), the orange curve represents the $(1_0,1_0,0,0,0)$ state, green $(1_1,1_0,0,0,0)\pm(1_0,1_1,0,0,0)$, red $(1_1,1_1,0,0,0)$, and purple $(0,0,1_0,0,0)+e^{i\theta}(0,0,0,1_0,0)+e^{2i\theta}(0,0,0,0,1_0)$, with $\theta=0,2\pi/3,4\pi/3$. The color scheme encodes $P_T$, and partially anti-symmetric states decoupled from the annealing, $\sum_{\ell_1,\ell_2,\dots,\ell_N} \langle n,\ell|\Psi\rangle=0$, are shown in gray. 
	}
	\label{fig:full_spectra}
\end{figure}

When a single MWICS is energetically dominant, the low-energy spectrum is described by the approximately factorizable  product states, cf.\ Eq.~\eqref{eq:product_state}. Consequently, the energies of the many-body modes are approximately additive, and in particular, the gap behaviors are inherited from the spectra and gaps of the factor states.

The paw graph $\mathcal{G}_{4-3}$ provides a representative example of this factorized structure, as evidenced by its spectrum in indent limit $L=20,d=3$ shown in Fig.~\ref{fig:full_spectra}(a). Its MICS consists of a single vertex $\{1\}$ and an edged pair $\{3,4\}$. The few relevant lowest-lying modes can be factorized as,
\begin{equation}	
	\label{eq:G4-3_product}
	\begin{tabular}{c|c}
		\hline
		Vertex config. & Quantum state \\
		\hline\hline
		$(1_{0/1},0,0,0)$ & $|\psi^1_{0/1}\rangle_1\otimes|\psi^0\rangle_2\otimes |K^2\rangle_{(3,4)}$ \\
		\hline
		$(0,1_0,0,0)$ & $|\psi^1_0\rangle_2\otimes\bigotimes_{n=1,3,4}|\psi^0\rangle_n$ \\
		\hline
		$(1_{0/1},0,1_0,0)$ & $|\psi^1_{0/1}\rangle_1\otimes|\psi^0\rangle_2\otimes$ \\
		$(1_{0/1},0,0,1_0)$ & $ \left(|\psi^1_0\rangle_3\otimes|\psi^0\rangle_4\pm |\psi^1_0\rangle_4\otimes|\psi^0\rangle_3\right)$ \\
		\hline
	\end{tabular}
\end{equation}
up to normalization.

In the early stage of the annealing, $\epsilon_{i}> t$, the ground state is given by $|\psi^1_0\rangle_1\otimes|\psi^0\rangle_2\otimes |K^2\rangle_{(3,4)}$. The instantaneous basis states are localized around $\bigotimes_{n=1}^{4} |n,1\rangle$, and can be interpreted as logical zero, $(0,0,0,0)$. The uniqueness of the quantum many-body ground state and the classical logical-zero configuration are consistent with each other. The first excited state is approximately four-fold degenerate, representing the four classical configurations $(1,0,0,0)$, $(0,1,0,0)$, $(0,0,1,0)$, and $(0,0,0,1)$. The energy difference between the ground state and these four states is roughly $\epsilon_{i}+t^2/\epsilon_{i}-2t$. As $\epsilon_{i}$ approaches $t$, the degeneracy is slowly lifted. The $(0,1_0,0,0)$ state obtains a relatively higher energy than the $(1_0,0,0,0)$ state, because the latter hosts a $|K^2\rangle_{3,4}$ bound state factor. Meanwhile, the symmetric and antisymmetric superposition states of $(1_0,0,1_0,0)$ and $(1_0,0,0,1_0)$ remain almost degenerate. 
The first gap minimum occurs near $\epsilon_i\approx t$. The quantum state undergoes a localization-to-delocalization crossover in the \nth{1} dimension, while keeping the common factors in the other dimensions unchanged. Consequently, their energy separation approximately follows the gap between the two lowest eigenstates, $|\psi^1_0\rangle$ and $|\psi^1_1\rangle$, of a one-dimensional chain. This product-state description reproduces the low-energy spectrum and eigenstates accurately, with comparatively more pronounced deviations near the gap minimum.

\begin{figure}[!t]
	\centering
	\includegraphics[width=\columnwidth]{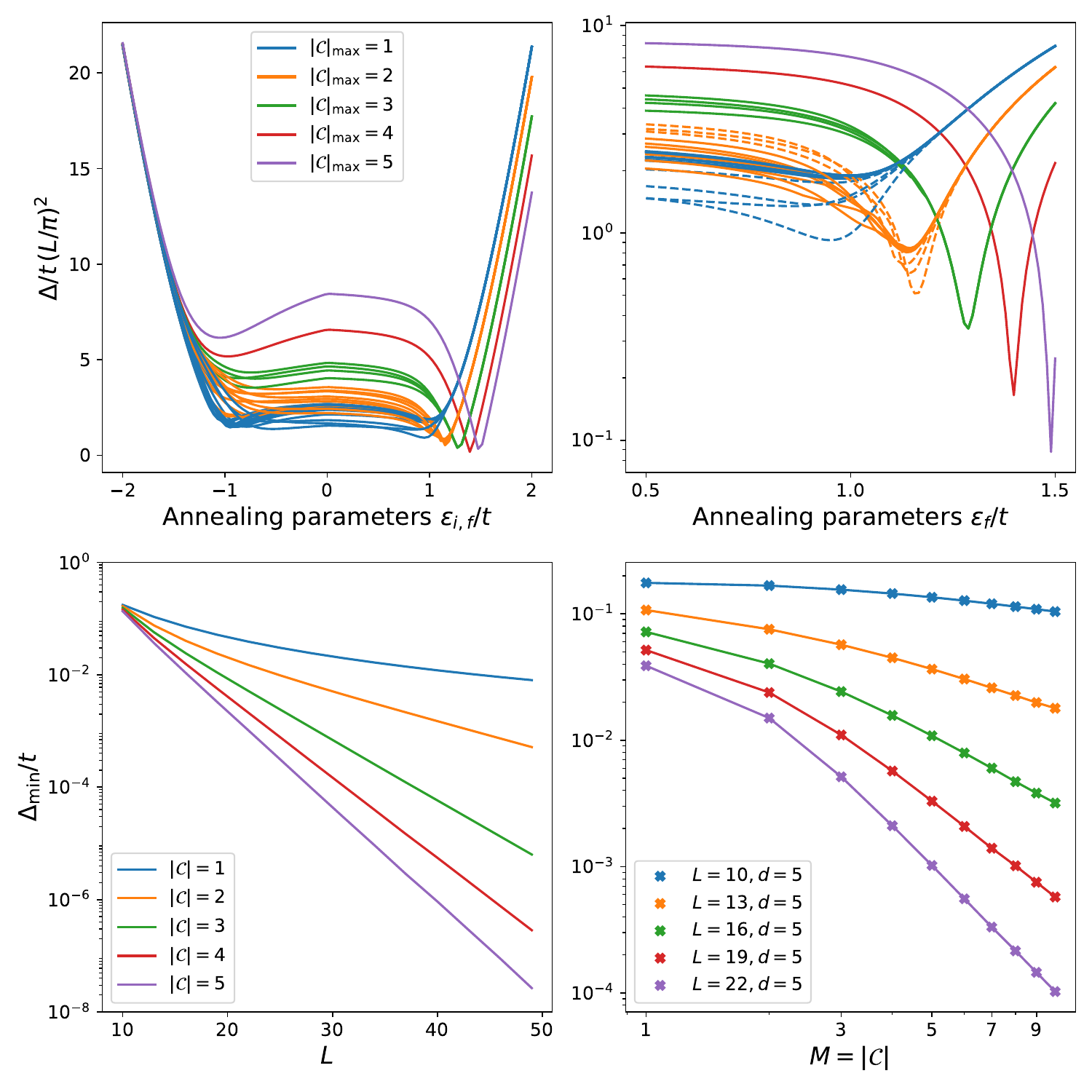}
	\caption{{\bf Scaling behaviors of gaps.} The gap in the second half of the annealing around $\epsilon_{f}\approx t$ shows spectral collapse associated to the largest clique in the MICS. These are shown (a) in the entire protocol in linear scale, and (b) in the vicinity of the minimum gap in logarithmic scale. This collapse is shown for all graphs with $N=2,3,4,5$ vertices, and for the implementation of $L=20,d=5$.
		Dashed curves indicate graphs, including $\mathcal{G}_{4-4}$, $\mathcal{G}_{4-5}$, $\mathcal{G}_{5-8}$, $\mathcal{G}_{5-11}$, $\mathcal{G}_{5-12}$, $\mathcal{G}_{5-19}$, $\mathcal{G}_{5-20}$. These graphs have high symmetry and subject to strong hybridization and/or binding effects, and their behaviors deviate from the generic scenarios.
		(c,d) Scaling behaviors of the minimum gap for complete graphs $\mathcal{K}_q$, computed using the recursive generation method. The minimum gap scales (c) exponentially with $L$, and (d) polynomially with $q$. These trends also apply to clique factors.
	}
	\label{fig:clique_univ}
\end{figure}

In the second half of the annealing, a second gap closure takes place around $\epsilon_{f}\approx t$. This corresponds to a transition within the union of MICS hypercuboid subspace, where the $(1,0,1,0)$ and $(1,0,0,1)$ states become energetically more favorable than the $(1,0,0,0)$ state. This transition concerns essentially the $|K^2\rangle_{(3,4)}$ bound state and the $\left(|\psi^1_0\rangle_3\otimes|\psi^0\rangle_4+|\psi^1_0\rangle_4\otimes|\psi^0\rangle_3\right)$ superposition state in the two-dimensional $\mathsf{L}$-shaped domain along the \nth{3} and \nth{4} dimensions. Particularly, these are the ground state and the second excited state, while the first excited state is decoupled from the annealing since its anti-symmetric structure between the \nth{3} and \nth{4} dimensions. 

Generically for factorizable quantum states, the behaviors of their gaps in the second half of the protocol are controlled by their cliques,
\begin{eqnarray}
	\label{eq:gap_universal}
	\Delta_\mathrm{min}(\mathcal{G};L,d) \approx \Delta_\mathrm{min}(M;L,d).
\end{eqnarray}
where $\Delta_\mathrm{min}$ refers to local minima. For factorizable quantum states, the quantum state already lies within the MICS hypercuboid subspace at the midpoint of the annealing. In the second half of the annealing, the factor clique-based bound state associated to each individual clique in $M\in\mathcal{M}$ transitions into superposition state independently. This dependence of gap behaviors on cliques is examined and confirmed for all small graphs with $N\leq 5$ vertices in Fig.~\ref{fig:clique_univ}(a,b). The minimum gap due to the cliques follows the behaviors of the corresponding complete graphs. For a few graphs with high symmetry, a deviation from the clique-based behaviors can be seen due to hybridization and binding.

We analyse the behaviors of the minimum gap for different cliques.
The low-energy spectrum associated to the complete graphs $\mathcal{K}_q$ can be solved using the \emph{recursive generation method}, as detailed in Appendix~\ref{sec:recursive_algorithm}. This method applies to a sequence of graphs which can be generated with a clear pattern, and assists us to probe the low-energy spectra for large $q$ and $L$ otherwise formidable with exact diagonalization.
The minimum gap scales exponentially with $L$ and polynomially with $q$ as $\Delta_\mathrm{min}\sim q^{-\alpha_q L}$, see Fig.~\ref{fig:clique_univ}(c,d). Particularly, the exponent coefficient goes from $\alpha_q\approx 0.20$ for $q=2$ to $\alpha\approx 0.25$ for $q=5$. This is consistent with the results presented in Ref.~\cite{bombieri25quantum}. See Appendix.~\ref{sec:deformation_MICS_appendix} for details.
Our analysis is further supported by $P_T$, Eq.~\eqref{eq:def_PT}, as illustrated by the color code in the spectrum, Fig.~\ref{fig:full_spectra}.

The paw graph $\mathcal{G}_{4-3}$ example illustrates the spectrum associated with a unique dominant MWICS. Consistent with our analysis for homogeneous boundary, hybridization effects between distinct MICS sectors are negligible. As a result, the low-energy problem admits an approximate reduction to the dynamics of its constituent factors. The gap behaviors is therefore controlled by the clique, leading to the clique-controlled behavior. As we clarify soon, this factorized description ceases to apply when several macroscopically distinct MWICS constructions become energetically competitive.

\subsection{Competing channels and hybridization induced gap closing}
\label{sec:competing}

We further investigate the case of competing ICSs, using graph $\mathcal{G}_{5-20}=\mathcal{K}_5\setminus e$ as an example. This graph has one edge removed from the $\mathcal{K}_5$ complete graph, and it has three competing ICSs: (i) a set of two vertices $\{1,2\}$ which is also the MIS, and (ii) two sets each consisting of a single 4-clique, which are second largest ICSs and referred to as suboptimal ICSs (subICSs). These relevant product states can be written as
\begin{equation}
	\begin{tabular}{c|c}
		\hline
		Vertex config. & Quantum state \\
		\hline\hline
		$(1_0,1_0,0,0,0)$ & $|\psi^1_0\rangle_1\otimes|\psi^1_0\rangle_2\otimes\bigotimes_{n=3,4,5}|\psi^0\rangle_n$ \\
		\hline
		$(1_1,1_0,0,0,0)$ & $(|\psi^1_0\rangle_1\otimes|\psi^1_1\rangle_2\pm |\psi^1_1\rangle_1\otimes|\psi^1_0\rangle_2)$ \\
		$(1_0,1_1,0,0,0)$ & $\otimes\bigotimes_{n=3,4,5}|\psi^0\rangle_n$ \\
		\hline
		$(1_1,1_1,0,0,0)$ & $|\psi^1_1\rangle_1\otimes|\psi^1_1\rangle_2\otimes\bigotimes_{n=3,4,5}|\psi^0\rangle_n$ \\
			\hline
		$(0,0,0,0,0)$ & $|K^4\rangle_{(1,3,4,5)}\otimes|\psi^0\rangle_{2}+|K^4\rangle_{(2,3,4,5)}\otimes|\psi^0\rangle_{1}$ \\
		\hline
		$(0,0,1_0,0,0)$ & $|\psi^0\rangle_1\otimes|\psi^0\rangle_2\otimes\big[|\psi^1_0\rangle_3\otimes|\psi^0\rangle_4\otimes|\psi^0\rangle_5$ \\
		$(0,0,0,1_0,0)$ & $+e^{i\theta}|\psi^0\rangle_3\otimes|\psi^1_0\rangle_4\otimes|\psi^0\rangle_5$ \\
		$(0,0,0,0,1_0)$ & $+e^{2i\theta}|\psi^0\rangle_3\otimes|\psi^0\rangle_4\otimes|\psi^1_0\rangle_5\big]/\sqrt{3}$ \\
		\hline
	\end{tabular} \nonumber \\
\end{equation}
The first three groups describe modes in the MIS subspace. The fourth row is the fully symmetric combination of the two $K^4$-based channels. The final group consists of three modes lying in the suboptimal ICS subspace, with $\theta=0$, $2\pi/3$, $4\pi/3$.

The automorphism of the graph $\mathcal{G}_{5-20}$ is isomorphic to the direct product of permutation groups $S_2$ and $S_3$, $\operatorname{Aut}(\mathcal{G}_{5-20})\simeq S_2\times S_3$. Here, $S_2$ permutes between vertices $1$ and $2$, and $S_3$ permutes among the vertices $3$, $4$, and $5$. These permutation symmetries are inherited by the quantum system. For each permutation symmetry $S_m$, there exist $m$ corresponding symmetry-related modes. Among them, a single fully symmetric mode is bright and coupled to the fully symmetric initial state, while the rest ${m-1}$ modes are dark and strictly inaccessible during the protocol. These include the antisymmetric superposition of $(1_1,1_0,0,0,0)$ and $(1_0,1_1,0,0,0)$ states, and the $\theta=2\pi/3$ and $\theta=4\pi/3$ superposition of $(0,0,1_0,0,0)$, $(0,0,0,1_0,0)$, and $(0,0,0,0,1_0)$ states. Consequently, only crossings between states in the accessible fully symmetric sector are dynamically relevant.

Within each orbit of \(m\) symmetry-related channels, one trivial-irrep combination is bright and the remaining \(m-1\) orthogonal combinations are dark.

In indent limit, the MICS has a larger weight. Correspondingly, the $(1_0,1_0,0,0,0)$ state is the ground state, and the low-energy quantum dynamics lies predominantly in the MIS hypercuboid. The spectrum for $L=20$, $d=3$ as shown in Fig.~\ref{fig:full_spectra}(b) supports this energetic picture, where the low-lying MICS-compatible quantum states are all characterized by $P_T\lessapprox 1$ as defined in Eq.~\eqref{eq:def_PT}. 

Although the spectrum largely matches a fully factorizable system, apparent deviation in the fully symmetric states can be observed at around $\epsilon_{i}\approx t$. In this regime, the $(1_0,1_0,0,0,0)$ state and the $(0,0,0,0,0)$ states are almost degenerate, leading to hybridization between them.
The hybridized spectrum preserves the overall branch structure inferred from the ansatz, but the anti-crossing in the vicinity of $\epsilon_{i}=t$ is substantially modified. Unlike the factor-controlled gap in Sec.~\ref{sec:factorizable}, this gap is not inherited from a single-chain spectrum, but rather controlled by the effective coupling between the two channels after taking their nonorthogonality into account.
This gap scales according to the effective coupling $\langle \mathrm{MICS}|H|\mathrm{subICS}\rangle$, which is suppressed by spatially small mode overlap.

Away from the indent limit, competition between the two ICSs affects the spectrum in a more pronounced manner. This is illustrated by the spectrum for  $L=20$, $d=10$ in Fig.~\ref{fig:full_spectra}(c). The low energy dynamics takes place predominantly in the subIS subspace, until the later stage of the annealing, $\epsilon_{f}\approx t$, where it transitions into the MIS subspace. This is consistent with the fact that the subICS is energetically more favorable than the MICS at homogeneous boundary.

The two geometries in Fig.~\ref{fig:full_spectra}(b,c) encode the same MIS problem, but favor different clique arrangements. At $d=3$, the two-vertex MICS is energetically favored, and its product state closely describes the midpoint ground state. At $d=10$, the single-clique subICS has a slightly larger weight, but the ground state has substantial overlap with both arrangements. Thus the clique weights rank the product configurations, and changing the chain geometry tunes the competition without changing the encoded problem.

\section{Dynamics towards target subspace}
\label{sec:dynamics}

The spectral structure of the blockaded-chain model motivates a diabatic annealing scheme that preserves the MIS target even though the dynamics leaves the ground state due to small energy gaps. The family of diamond-chain graphs provide a concrete playground for investigating the effectiveness of the scheme.

\subsection{Diamond chain graphs as a recursive family of competing sectors}
\label{sec:diamond}

\begin{figure}[!t]
	\centering
	\includegraphics[width=\columnwidth]{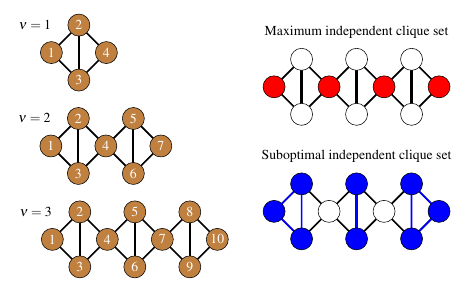}
	\caption{{\bf Diamond-chain graphs.} (a-c) The first three diamond-chain graphs, $\mathcal{D}_\nu$, $\nu=1,2,3$. (d,e) The maximum and suboptimal independent clique sets for $\mathcal{D}_3$, each possessing total weight $W_\mathrm{MICS}=4W_1$ and $W_\mathrm{subICS}=W_2+2W_3$, respectively. The suboptimal solution is energetically more favorable when $W_\mathrm{subICS}>W_\mathrm{MICS}$.}
	\label{fig:diamond_chain}
\end{figure}

\begin{figure}[!t]
	\centering
	\includegraphics[width=\columnwidth]{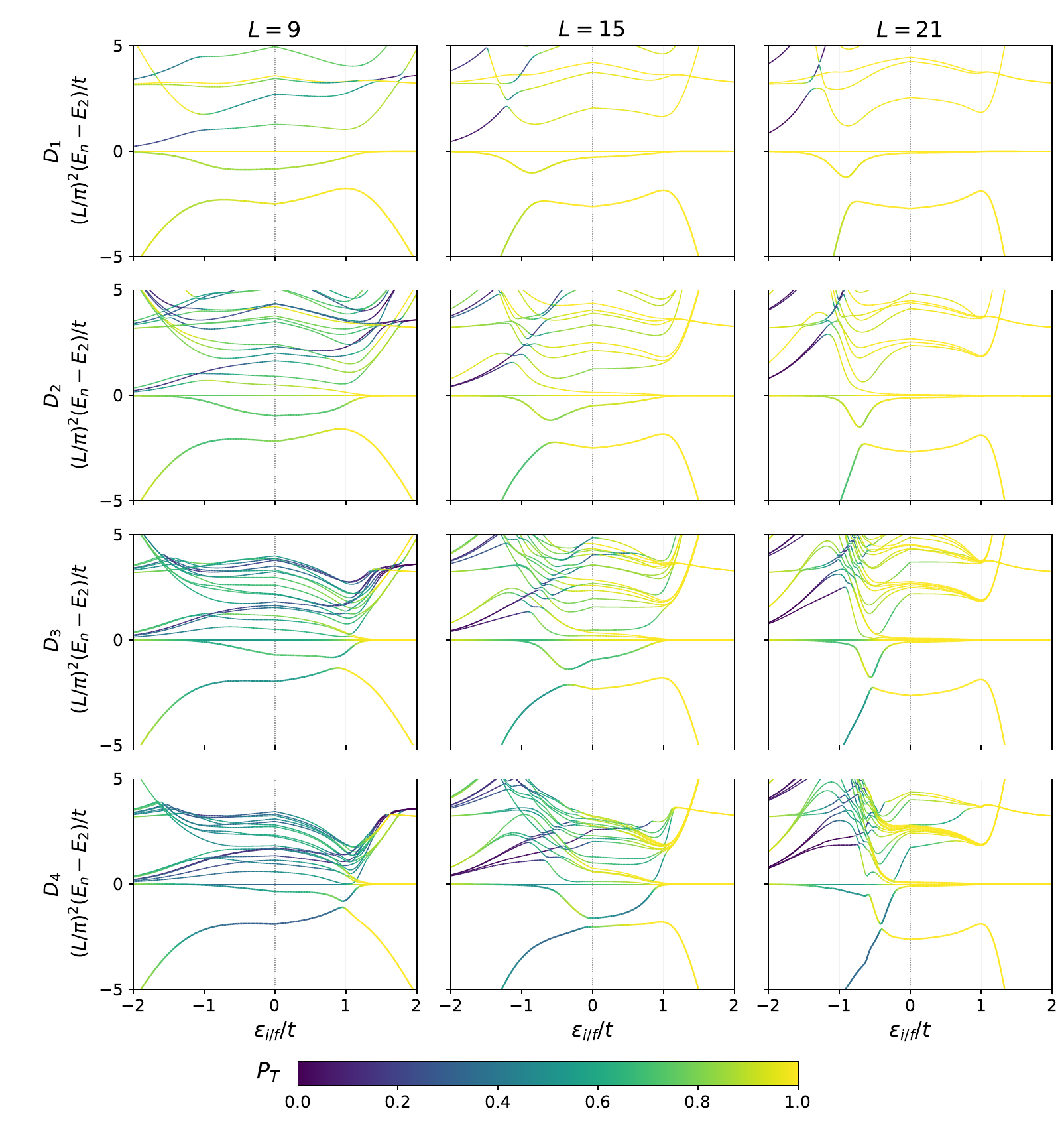}
	\caption{{\bf Diamond chain spectra.} 	The spectra of the lowest 100 modes for the $\mathcal{D}_1$ to $\mathcal{D}_4$ diamond chain graphs for different $L=9,15,21$ with fixed $d=3$. The color encodes whether the branch lies within the correct MICS subspace. The position of the minimum gap shifts towards an earlier stage as the indent limit is approached.
	}
	\label{fig:diamond_chain_spectra}
\end{figure}

The \emph{diamond chain graphs}, $\mathcal{D}_\nu$, provide a paradigmatic example for investigating competition between different ISs in quantum annealing. As shown in Fig.~\ref{fig:diamond_chain}(a), $\mathcal{D}_\nu$ is constructed by an initial vertex attached by a chain of $\nu$ 3-cyclic graphs $\mathcal{K}_3$, which adds up to $N=3\nu+1$ vertices in total.
The diamond-chain family provides a controlled setting for studying competition between independent set sectors. Each graph $\mathcal{D}_\nu$ has a unique MIS with $\nu+1$ vertices, together with $O(2^\nu)$ subISs with $\nu$ vertices. This exponential proliferation of almost-optimal configurations creates a competing landscape that can hinder annealing dynamics.

Meanwhile, the performance of quantum annealing depends on the weights of the cliques. For $\nu\ge2$, the MICS of the graph is exactly the MIS, with $(\nu+1)$ $\mathcal{K}^1$; while the subICS consists of two $\mathcal{K}^3$ and $(\nu-2)$ $\mathcal{K}^2$, compatible to the subIS. For example, the two competing ICSs for $\mathcal{D}_3$ are shown in Fig.~\ref{fig:diamond_chain}(d) and (e). When $2W_3+(\nu-2)W_2-(\nu+1)W_1<0$, the MICS has the largest weight.
The diamond chain graphs follow a clear pattern. Therefore, their associated low-energy spectra and dynamics can be computed with the recursive generation method, allowing us to probe a larger range of $N$ and $L$, see Appendix~\ref{sec:recursive_algorithm} for details. The spectra shown in Fig.~\ref{fig:diamond_chain_spectra} substantiate the relation between the quantum dynamics and the weights of the ICSs. Given fixed $L$ and $d$, the spectrum associated to $\mathcal{D}_\nu$ is characterized by a single gap closure, where the system transitions from the subICS state to the MICS state. 

The position of this gap is shifted by the shape ratio. Away from the indent limit, $W_2>\frac{3}{2}W_1$, the subICS has a larger weight. As the indent limit is approached, the weights of different cliques become more uniform, and the MICS thus becomes eventually more favorable. Correspondingly, the quantum state at the midpoint of the protocol achieves the subICS configuration away from indent limit, and achieves the MICS configuration in indent limit. The gap minimum shifts towards earlier stage of the annealing with decreasing ratio of $d/L$, for a given diamond-chain graphs $\mathcal{D}_\nu$, see Fig.~\ref{fig:diamond_chain_spectra}. Concomitantly, increasing number of low-energy modes lies in the MICS subspace, and all the relevant low-energy branches becomes MICS compatible for large $L=24$. The spectrum structure crosses over between the scenarios illustrated in Fig.~\ref{fig:full_spectra}(b) and (c). Particularly, the minimum possesses a roughly exponential scaling behaviors, similar to the spinor implementation~\cite{schiffer24circ,lukin2024quantum}, while a more precise estimate of the scaling is not yet achievable.

\subsection{Target-preserving diabatic annealing scheme}

\begin{figure}[!t]
	\centering
	\includegraphics[width=\columnwidth]{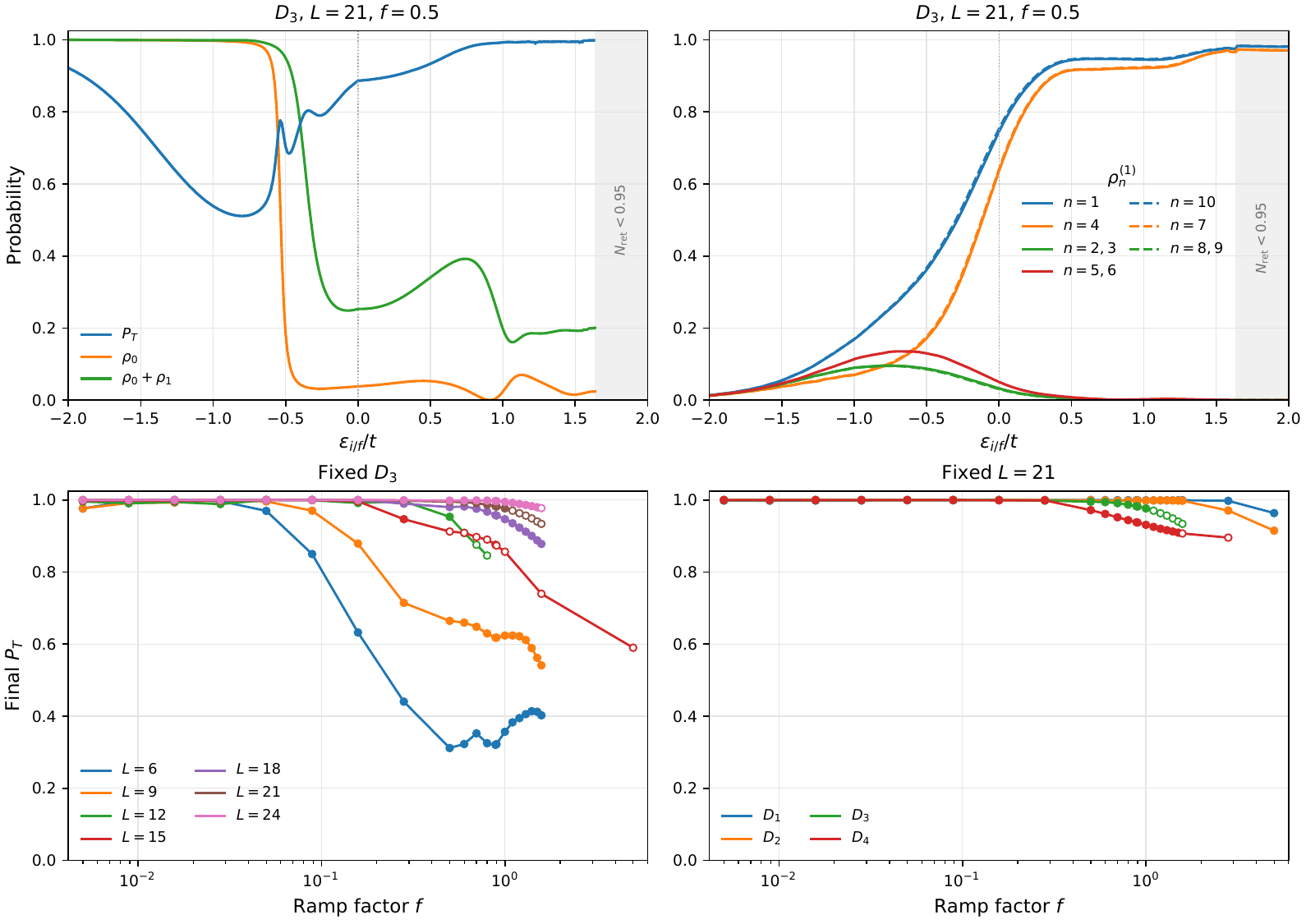}
	\caption{{\bf Target-preserving dynamics in diamond-chain graphs.} The system is prepared in the ground state at $\epsilon_{i}=2t$, and evolved along the logical protocol, cf. Fig.~\ref{fig:protocol}, until $\epsilon_{f}=2t$.
	(a,b) The dynamics of the quantum state on $\mathcal{D}_3$ implemented with $L=21,d=3$ and ramp rate $f=0.5$. (a) The quantum state leaves the ground state (orange curve, $\rho_0$) and first excited state (green curve, $\rho_0+\rho_1$), but eventually reaches the target subspace (blue curve, $P_T$). (b) The instantaneous experimentally relevant density readout $\rho^{(1)}_n$ along different vertices. Pairs like $n=1$ and $n=10$ are symmetric in the physical system, but asymmetry arises due to numerical truncation. They are shown by solid and dashed lines, respectively, of the same color. The shaded regime indicates that the mode truncation is numerically not fully converged.
	(c,d) The final dephased weight in the target subspace, $P_T$, cf. Eq.~\eqref{eq:def_PT}, with various ramp rates $f$, is shown for (c) varying $L$ for $\mathcal{D}_3$, and (d) for $\mathcal{D}_1$ through $\mathcal{D}_4$ at fixed $L=21$. A mode truncation at $M=100$ is used, and the hollow points indicate that the corresponding numerical convergence is not fully converged. 
	}
	\label{fig:diamond_chain_dynamics}
\end{figure}

The spectra shown in Fig.~\ref{fig:diamond_chain_spectra} motivate us to use the logical annealing protocol diabatically, c.f. Sec.~\ref{sec:protocol}, for recovering quantum states representing the MIS solution. Away from the indent limit, the suboptimal ICS state is favored at the homogeneous midpoint, and the lowest branch acquires MIS character only after a narrow gap in the second half of the anneal. This spectral feature constrains the feasible annealing rate. Near the indent limit, the MICS state is favored energetically at the midpoint, together with several low-energy branches that carry substantial weight in the MIS sector. This structure suggests that diabatic transitions can deplete the instantaneous ground-state population while allowing substantial target weight to be recovered by the end of the anneal.

The target-preserving feature of the diabatic annealing can be verified with dynamical evolution using the recursive generation method, see Appendix~\ref{sec:recursive_algorithm_dynamics}. We initialize the system at the instantaneous ground state at $\epsilon_{i}=2t$, and evolve it under a normalized ramping rate of $f = \dot{\epsilon}_{i/f} (L/\pi)^2/t^2$ ($\hbar=1$) with the logical protocol, and compute the occupation of the MICS subspace, $P_T$  defined in Eq.~\eqref{eq:def_PT}, by the end of the annealing at $\epsilon_{f}=2t$.

The $\mathcal{D}_3$ model implemented in indent limit $L=21,d=3$ hosts characteristic dynamics compatible to its spectrum, see Fig.~\ref{fig:diamond_chain_dynamics}(a). Along with the several small gaps in the spectrum, the weight of the evolving quantum state in the ground state, $\rho_0$, and first excited state, $\rho_1$, drops substantially in the first half of the protocol, close to $\epsilon_{i}\approx0.5t$. However, the quantum state does not leave the target subspace indefinitely. In fact, the instantaneous weight in the target subspace drops initially. This detour inherits from the dressing of the ground state by $|\mathrm{subICS}\rangle$ states. As the annealing approaches the homogeneous midpoint, the dressing effects become weak, and the state returns to the target subspace. As the annealing further proceeds, the weight of the state in excited states retains, but it remains substantially in the target subspace. We remark that the system does not need to be restrained in the ground state throughout the annealing to preserve its target by the end of the annealing. 

The preservation of the target can be experimentally measured by the partial density distribution along different dimensions,
\begin{eqnarray}
	\rho^{(1)}_n = \sum_{\ell_n=d+1}^{L} \rho_n(\ell_n),
\end{eqnarray}
 c.f.\ Eq.~\eqref{eq:def_rho_n} and see Appendix.~\ref{sec:chain_from_network} for details. As shown in Fig.~\ref{fig:diamond_chain_dynamics}(b), the targeted vertices, $n=1,4,7,10$, has increasing readouts along the annealing, while the irrelevant vertices has eventual vanishing readouts. This confirms that target weight $P_T$ is compatible with experimentally measured solution readouts.

We further investigate the scaling of the ramp rate while changing the graph order $\nu$ and the system size $L$.
Given the graph $\mathcal{D}_3$, we increase $L$ from $L=6$ to $L=24$ at fixed $d=3$, see Fig.~\ref{fig:diamond_chain_dynamics}(c). Far from the indent limit, i.e., for $L=6,9,12$, it takes exponentially slower ramp rates to reach the target subspace $P_T\approx1$. This is consistent with the spectral structure of the system. In this case, the subICS is energetically more favorable at mid-protocol, and the system needs to go through the exponentially closing gap in the second half of the annealing, for entering the MICS subspace. At larger $L=18,21,24$, the system achieves $P_T\approx1$ at roughly $f=1$, and the required ramp rate is much less sensitive to the change in $L$. This is consistent with the fact that the MICS subspace is always energetically more favorable in indent limit.

The annealing performance among different graphs $\mathcal{D}_{1,2,3,4}$ at fixed $L=21$, $d=3$ is shown in Fig.~\ref{fig:diamond_chain_dynamics}(d). The majority of the system weight enters the target subspace $P_T\gtrsim0.9$ for all probed ramp rates. There is further a drop of $P_T$ from $P_T=1$ to $P_T\approx0.9$ at around $f\approx1$, and this drop takes place at slightly faster $f$ for larger graph order $\nu$. This indicates the system achieves a fixed, substantial target weight through diabatic evolution within a ramp time scaling as $O(L^2)$, even in the presence of narrow intra-subspace gaps.

These results distinguish adiabatic ground-state preparation from target-compatible diabatic evolution. Although the ramp substantially depletes the lowest instantaneous modes, it recovers the target weight, while suppressing unwanted logical occupations. The resulting dynamics therefore involves a broader target manifold, rather than the single ground state. 

The diabatic annealing implemented on diamond-chain graphs thus provides an example, where engineering the shape ratio of the system yields a spectral structure. This structure facilitates reaching the target MICS subspace associated to the correct classical solution. This scheme satisfies the unbiasedness criteria specified in Sec.~\ref{sec:protocol}. For sufficiently high-order diamond chains, the competing subICS becomes energetically favorable even in the indent limit, so the energetic rationale supporting the present protocol no longer applies. Since $\lim_{d/L\to0} W_2 \approx 1.071 > 1$, the subICS has a larger weight than the MICS for $\mathcal{D}_{\nu}$ with $\nu \ge 15$ $(N\ge46)$ for any choice of $d/L$. By further comparison, the diabatic scheme of Ref.~\cite{lukin2024quantum} employs an optimized detuning quench adapted to optimized quench parameters, to populate an excited branch that eventually connects to the final ground state.

\section{Discussion and conclusions}
\label{sec:conclusions}

We discuss the atomic overhead required by an experimentally relevant implementation. The standard Rydberg network construction for an $N$-vertex graph requires around $4N^2$ atoms~\cite{nguyen23quantum}, arranged into $N$ wires of length $O(N)$. To approach the indent regime, we extend the logical-one end of each wire, increasing the ratio $\xi=L/d$ in accordance to Fig.~\ref{fig:entropy_fidelity}. This requires $O((\xi-1)N)$ additional atoms per wire, giving a total atom count of $O(\xi N^2)$. Thus, for a fixed elongation ratio, the atomic overhead remains quadratic in $N$. 

When the target-compatible sector is energetically favorable, diabatic evolution populates excited states without necessarily compromising the final target weight. For a fixed normalized ramp factor $f$, the protocol requires an annealing time scaling as $O(L^2)$, where we attain high final target weight. In Rydberg network implementation $L=\xi N$, the run-time overhead is therefore $O(N^2)$.

Additional avoided crossings arise from transitions involving bound states. For the clique examples studied at fixed $d$, the corresponding minimum gaps decrease approximately exponentially with $L$. Strategies for mitigating these spectral bottlenecks require further study. Nevertheless, failure to traverse such crossings adiabatically does not immediately invalidate the computation results in the final state. Within the factorized MICS description, retaining a bound state suppresses logical-one readouts within the corresponding vertices, without altering the readouts of the other factors. Consequently, the reported logical-one vertices form a partial independent set that can be reliably extended into an MIS. The partial solution from the quantum readout can significantly reduce the remaining classical optimization problem.

We propose further improvement to the encoding scheme that potentially alleviates the annealing bottleneck in a fundamental manner. The underlying mechanism of annealing performance lies in energetic competition, where suboptimal solutions potentially become more favorable due to quantum binding effects in the logical-zero sector $[0,d]^{\otimes N}$. This observation motivates us to elevate the energy in that particular region by implementing an extra potential $H_\mathrm{core} = \epsilon_0 \sum_{n=1}^{N} \sum_{\ell=1}^{d} |n,\ell\rangle\langle n,\ell|$, with $\epsilon_0>0$. The effectiveness of this extra potential for alleviating the bottleneck worths future study.

To summarize, we have investigated the low-energy spectrum and annealing dynamics of blockaded-chain models, which encodes the maximum independent set problems. These systems host competing many-body configurations whose approximate energetics can be characterized by physically derived clique and binding motif weights. Engineering the Hilbert-space geometry controls the relative energies of target-compatible and competing configurations. From the graph theoretical perspective, it engineers a cost function which relies on the global configuration. In contrast, standard classical annealing can feasibly impose cost functions based on local graph structures, which is less sensitive the global features of the graph. Diabatic protocols exploits the resulting spectral structure, during which the state leaves the instantaneous ground state yet recovers high final target-subspace weight. Our work indicates that solving a class of combinatorial optimization problems using quantum annealing in neutral atom networks can potentially improve annealing performance. The effectiveness and robustness of this approach in larger and more complex networks remain to be further explored.

Beyond optimization problems, neutral atom networks offer a flexible setting for investigating how engineered Hilbert-space geometry shapes localization and transport. This setup has thus demonstrated its flexibility in investigating  many-body localization, and can facilitate future studies on geometric effects in Hilbert-space fragmentation~\cite{moudgalya22quantum,moudgalya22hilbert,li23hilbert}, and quantum thermalization~\cite{deutsch18eigenstate,ueda20quantum}.

\begin{acknowledgments}
	I would like to thank Hannes Pichler for initial conceptualization, Zhongda Zeng and Roberto Tricarico for in-depth disucssions, and Yuanyuan Zhao for inspirational discussion on the recursive generation algorithm.
\end{acknowledgments}

\begin{appendices}

\section{List of graphs}
\label{sec:graph_list}
For reference, we list all the small graphs with $N\le5$ vertices~\cite{read05atlas} in Table~\ref{tab:graph_list}, together with their maximum independent sets and maximum independent clique sets.

\begin{table}[!t]
	\centering
	\caption{{\bf Atlas list of small graphs, and their maximum independent vertex and clique sets.} 
	}
	\label{tab:graph_list}
	\includegraphics[width=\columnwidth]{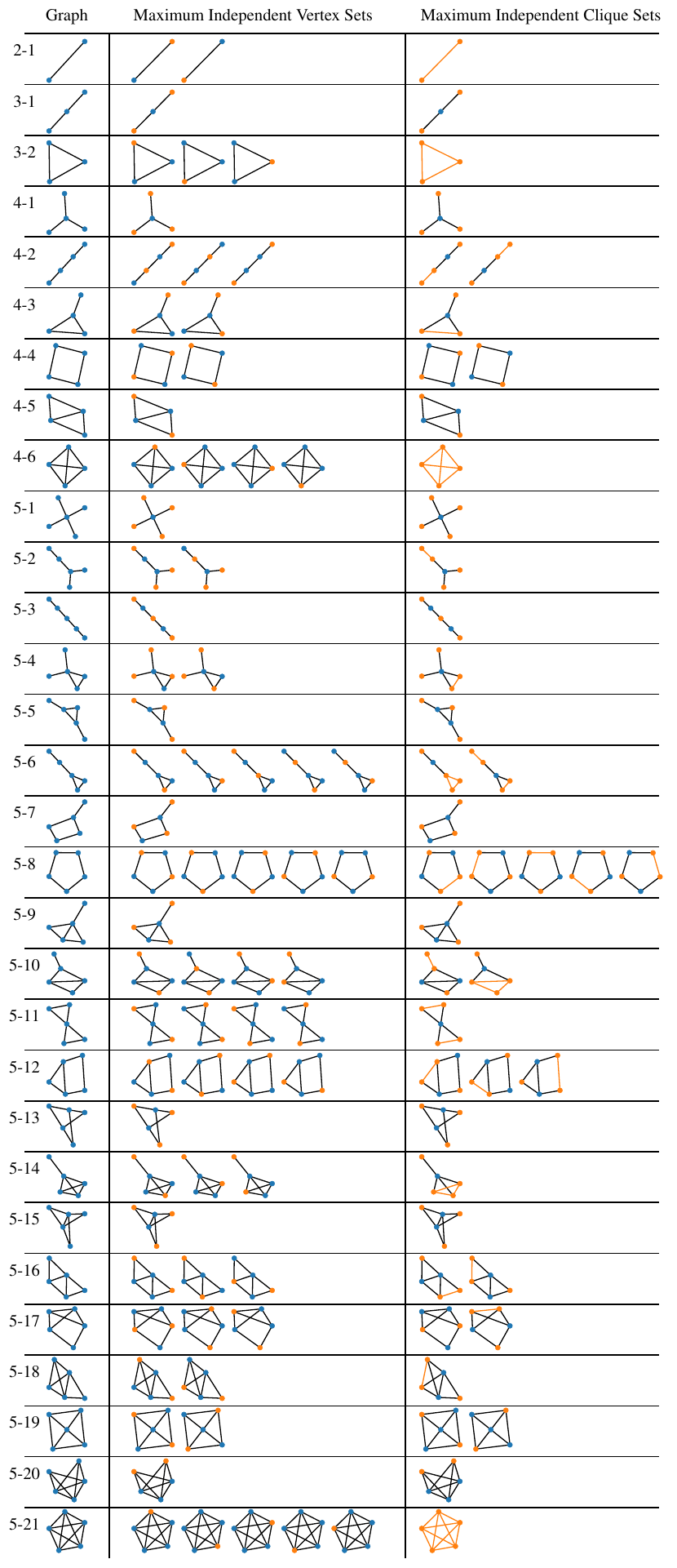}
\end{table}

\section{Blockaded-chain model as effective low-energy model for Rydberg networks}
\label{sec:chain_from_network}

We review the scheme using Rydberg network to encode the maximum independent set (MIS) problem~\cite{nguyen23quantum}, and then derive the blockaded chain model that we substantially investigate throughout this work.

\begin{figure}[!t]
	\centering
	\includegraphics[width=\columnwidth]{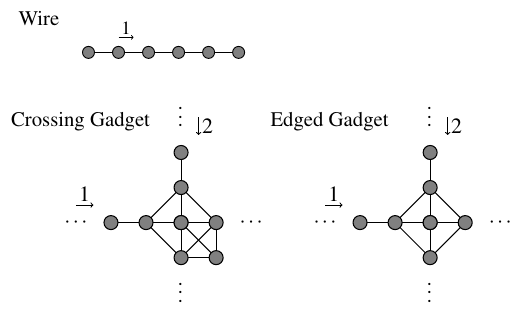}
	\caption{(a) A Rydberg wire of 6 atoms encoding the quantum degree of freedom associated to a vertex. (b) A crossing gadget not imposing constraint between the two vertices. (c) An edged gadget forbidding simultaneous occupation of the logical-one state of the two vertices. The arrow points to right, indicate that the logical-zero configuration is $|\uparrow\downarrow\uparrow\downarrow\uparrow\downarrow\rangle$, while the logical-one configuration is $|\downarrow\uparrow\downarrow\uparrow\downarrow\uparrow\rangle$.
	}
	\label{fig:schematics_network}
\end{figure}

In this encoding scheme, an array of neutral atoms subject to nearest-neighbor Rydberg interaction, also referred to as a \emph{Rydberg wire}, encodes the occupation of a graph vertex, see Fig.~\ref{fig:schematics_network}(a). In each wire, an even number of atoms are subject to uniform Rabi frequency $\Omega$ and detunings $\Delta$, such that each of the two $Z_2$ configurations represents the logical-zero (unoccupied vertex) and logical-one (occupied vertex), respectively. Multiple wires are then arranged into a two-dimensional network, such that any pair of wires crosses with each other. The crossings are always placed at $(2d^{<m,n>}_m-1)$-th and $(2d^{<m,n>}_n-1)$-th atoms of the wires.
At the crossings, the atoms are arranged into patterns called \emph{gadgets}. The crossing gadget does not implement additional constraint to the configurations of the two corresponding wires, and thus encode a pair of vertices not connected by an edge, see Fig.~\ref{fig:schematics_network}(b). In contrast, the edge gadget imposes an additional constraint that the two wires cannot occupy the logical-one mode simultaneously, and thus encodes a pair of vertices connected by an edge, see Fig.~\ref{fig:schematics_network}(c).

The Rydberg network can thus be formally described by a PXP Hamiltonian,
\begin{eqnarray}
	H &=& \sum_{i} \Pi^\mathrm{network} (\Omega \sigma_i^x - \Delta \sigma_i^z) \Pi^\mathrm{network} \\
	\Pi^\mathrm{network} &=& \prod_{n=1}^{N} \Pi^\mathrm{wire}_n \times \prod_{m<n} \Pi^\mathrm{gadget}_{m,n}
\end{eqnarray}
where $i$ sums over all neutral atoms in the network, $\sigma_i^{x/z}$ are Pauli matrices acting on the atomic Rydberg level $|\uparrow\rangle$ and ground level $|\downarrow\rangle$. The product in $\Pi^\mathrm{wire}$ run over all wires, with $\Pi^\mathrm{wire}_n=\prod_{i=1}^{2L-1}\Pi_{i,j}$ run through all neighboring atoms; while the product in $\Pi^\mathrm{gadget}$ run over all pairs of wires, with $\Pi^\mathrm{gadget}_{m,n}$ following the blockade rule established by Fig.~\ref{fig:schematics_network}(b,c). Here, $\Pi_{i,j} = 1_i\otimes1_j - |\uparrow_i\uparrow_j\rangle\langle \uparrow_i\uparrow_j|$ projects out the simultaneous occupation of Rydberg levels of two blockaded atoms.

When $\Delta\gg \Omega$, we can work in the domain wall limit~\cite{bombieri25quantum}. The domain wall basis states are the highly degenerate ground states of the system when taking $\Omega=0$. For a single wire of $2(L-1)=8$ atoms, these basis states are
\begin{align}
	& |\uparrow\downarrow\uparrow\downarrow\uparrow\downarrow\uparrow\downarrow\rangle, 
	|\uparrow\downarrow\uparrow\downarrow\uparrow\downarrow\downarrow\uparrow\rangle, 
	|\uparrow\downarrow\uparrow\downarrow\downarrow\uparrow\downarrow\uparrow\rangle, \nonumber\\
	& |\uparrow\downarrow\downarrow\uparrow\downarrow\uparrow\downarrow\uparrow\rangle, 
	|\downarrow\uparrow\downarrow\uparrow\downarrow\uparrow\downarrow\uparrow\rangle.
\end{align}
We label these states by $\ell_n=1,2,\dots,L$, and identify $\ell_1$ and $\ell_L$ to logical-zero and logical-one edges, respectively. When all gadgets are crossing gadgets and thus impose no constraint, they consitute $L^N$ domain-wall states, which can be labeled by coordinates $(\ell_1,\ell_2,\dots,\ell_N)$, $\ell_n=1,2,\dots,L$. In the presence of edged gadgets, the additional constraint thus forbids selected domain-wall states. Specifically, any states with coordinates $\ell_m>d^{<m,n>}_m$, $\ell_n>d^{<m,n>}_n$ become illegitimate.

Once $\Omega$ is turned on, it provides a second-order perturbative channel, where one atom transitions from its Rydberg state to its ground state, while another atom transitions from its ground state to its Rydberg state. This perturbative process takes place for two neighboring atoms which are subject to the Rydberg blockade $\Pi$. In the domain-wall basis states, this effectively induces a hopping of domain wall with strength $t=\Omega^2/2\Delta$. Along the wires, this hopping involves neighboring states, leading to $h_n$ as defined in Eq.~\eqref{eq:intact_cube}. In the vicinity of the gadgets, it induces local terms $h^\mathrm{gadget} = |d^{<m,n>}_m,d^{<m,n>}_n+1\rangle\langle d^{<m,n>}_m+1,d^{<m,n>}_n| + |d^{<m,n>}_m,d^{<m,n>}_n\rangle\langle d^{<m,n>}_m+1,d^{<m,n>}_n+1| + \mathrm{h.c.}$.
As an example, for the realistic implementation shown in Fig.~\ref{fig:illustration}(e), its effective domain-wall Hamiltonian $H^\mathrm{dw} = \Pi^\mathrm{dw} (h_1+h_2+h_3+h^\mathrm{gadget}_{12}+h^\mathrm{gadget}_{23}) \Pi^\mathrm{dw}$ is illustrated in Fig.~\ref{fig:illustration}(f).

To achieve the final symmetric form of the blockaded-chain model as presented in Eq.~\eqref{eq:model} and investigated throughout this work, we further simplify the domain-wall model by omitting the gadget hoppings as detailed in Ref.~\cite{bombieri25quantum}, and the gadget structure by taking the idealized uniform gadget positioning limit, $d^{<m,n>}_m=d$. This simplification is illustrated by Fig.~\ref{fig:illustration}(g).

A complementary and experimentally relevant measurement for $P_T$, Eq.~\eqref{eq:def_PT}, is provided by the coordinate-resolved density $\rho_n(\ell_n)$, Eq.~\eqref{eq:def_rho_n}. We can quantify whether the quantum state lies within the logical-one region by its single-body probability, $\rho^{(1)}_n = \sum_{\ell_n=d+1}^{L} \rho_n(\ell_n)$.
This quantity is experimentally accessible with Rydberg arrays. In the Rydberg array, this probability translates into the occupation of the Rydberg level of the $(2d)$th atoms, $(\langle \sigma^z_{n,2d}\rangle+1)/2$, which is easily accessible in experimental measurements.

\section{Dynamics and mapping in hypercube limit}
\label{sec:hypercube_two_mode}

\begin{figure}[!t]
	\centering
	\includegraphics[width=\columnwidth]{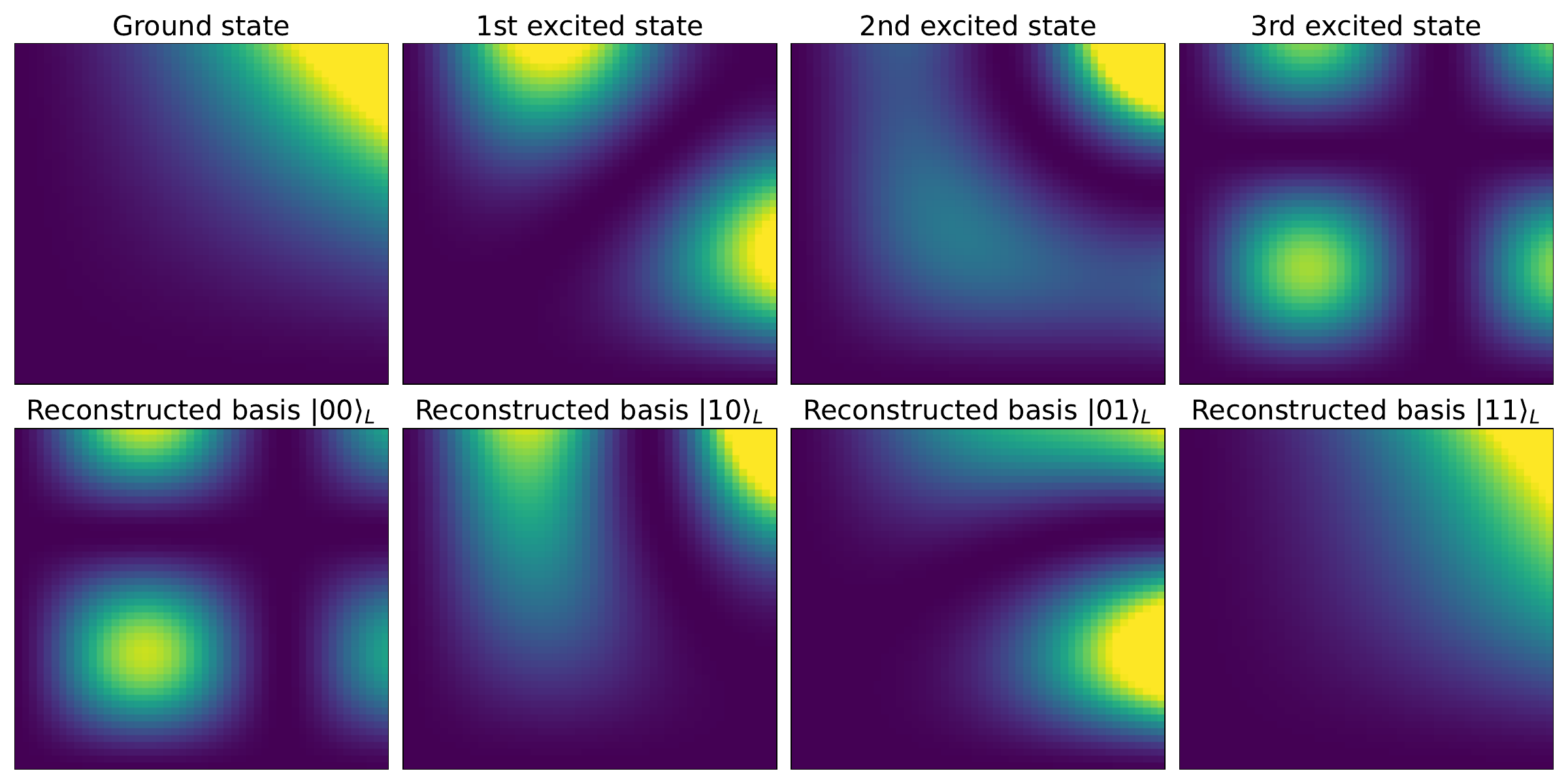}
	\caption{Low-energy eigenstates and reconstructed basis in a two-dimensional model without blockade. Parameters are taken as $L=50$, $\epsilon_i=0$, $\epsilon_f=t$.
	}
	\label{fig:mode_expansion_CwoE}
\end{figure}

\begin{figure}[!t]
	\centering
	\includegraphics[width=\columnwidth]{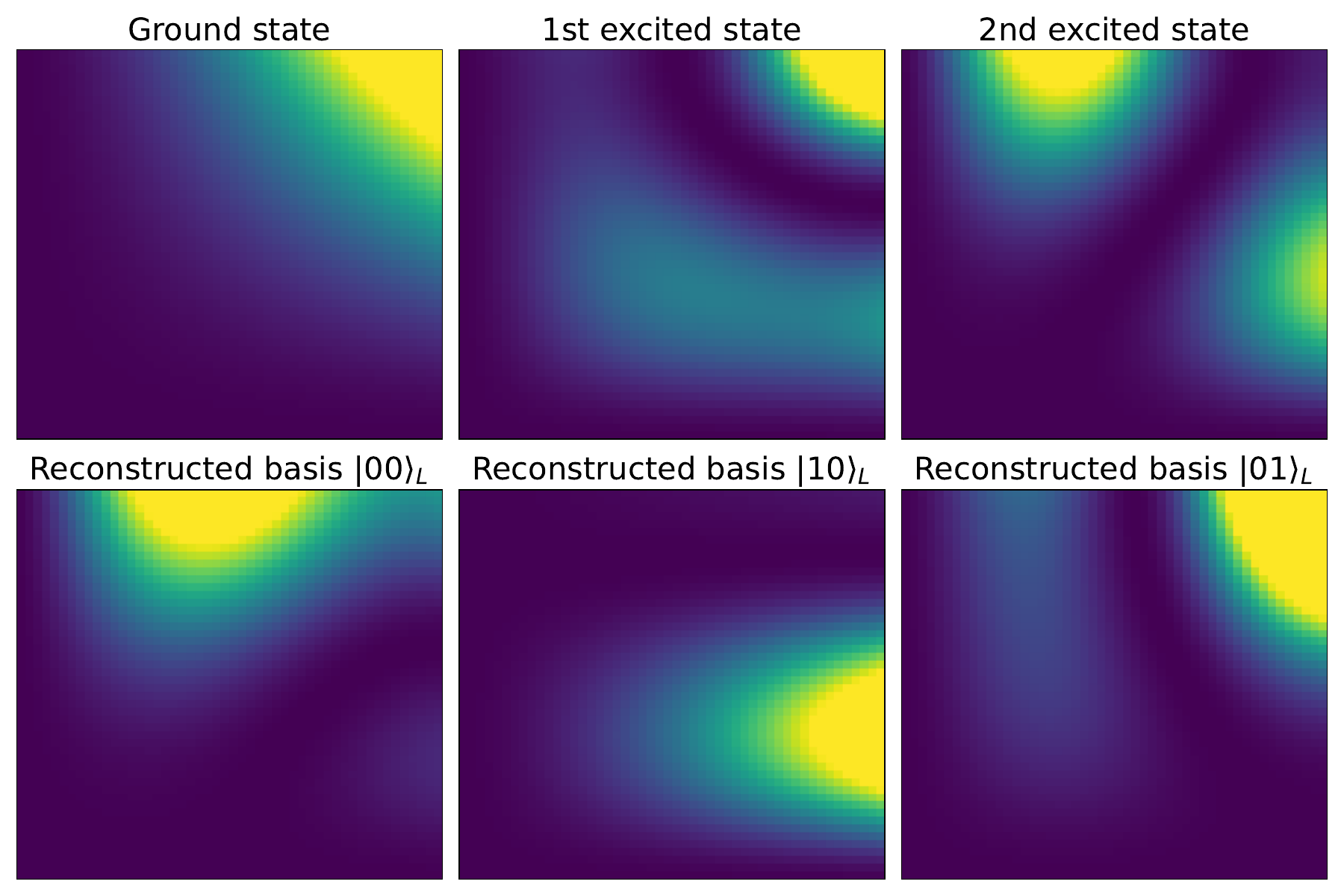}
	\caption{Low-energy eigenstates and reconstructed basis in a two-dimensional model with blockade. Parameters are taken in hypercube limit, $L=50,d=49$, $\epsilon_i=0$, $\epsilon_f=t$.
	}
	\label{fig:mode_expansion_CWE}
\end{figure}

In the hypercube limit, $d=L-1$, a graph edge excludes only the simultaneous occupation of the final sites of the two corresponding chains. Although the excluded fraction of the configuration space is small, the constraint can affect low-energy modes localized near that boundary. To achieve an effective low energy description, we first solve the finite one-dimensional problem and identify its crossover energy scale. This effective energetic truncation of the blockaded-chain model leads to a two-mode variational description with the connectivity of the PXP implementation.

\subsubsection{Finite-chain eigenstates}

We solve the low-energy modes of a single chain. The one-dimensional Hamiltonian in Eq.~\eqref{eq:intact_cube_onebody} leads to the bulk equation $E\psi_\ell=-t(\psi_{\ell-1}+\psi_{\ell+1})$ and boundary conditions $\psi_2 = (-E/t-\epsilon_i/t) \psi_1$, $\psi_{L-1} = (-E/t-\epsilon_f/t) \psi_L$. Along the logical protocol, see Fig.~\ref{fig:protocol}, the local potential on one edge vanishes. We take $\epsilon_i\geq 0$ and $\epsilon_f=0$ without loss of generality.

We are most interested in the regime $\epsilon_i\gtrsim t$, where the ground mode crosses over from a standing-wave mode to a exponentially decaying mode.

For $\epsilon_{i}$ much larger than $t$, the ground mode is localized on the edge and decays exponentially along the chain,
\begin{eqnarray}
	\psi^0_{\ell}\propto\sinh[\kappa(L+1-\ell)]
\end{eqnarray}
and possesses an energy below the continuum band, $E<-2t$,
\begin{eqnarray}
	E = -2t \cosh\kappa
\end{eqnarray}
where the decay rate $\kappa$ is fixed by boundary condition on $\ell=0$ edge as $\sinh(\kappa(L+1))/\sinh(\kappa L)=\epsilon_{i}/t$.
When the length of the chain is much longer than the decay length, $\kappa L \gg 1$, the wavefunction and the energy obtain explicit forms 
\begin{eqnarray}
\psi_{\ell} = \sqrt{1-t^2/\epsilon_i^2} (\epsilon_i/t)^{-\ell+1}
\end{eqnarray}
and
\begin{eqnarray}
	E/t = - \epsilon_{i}/t - t/\epsilon_{i},
\end{eqnarray}
respectively. In the regime $1/L\ll r \ll 1$, $r=\epsilon_{i}/t-1$, the energy can be further reduced to
\begin{eqnarray}
	E/t \approx -2 - (\epsilon_{i}/t-1)^2.
\end{eqnarray}

When $\epsilon_{i}$ approaches $t$, the ground mode becomes a standing-wave mode. Its wavefunction and energy are
\begin{eqnarray}
	\psi^0_{\ell}\propto\sin[k(L+1-\ell)]
\end{eqnarray}
and
\begin{eqnarray}
	E = -2t \cos k,
\end{eqnarray}
where the wave vector is given by $\sin(k(L+1))/\sin(kL)=\epsilon_{i}/t$.
In the regime $r\ll 1/L$, the wave vector can be approximated by
\begin{eqnarray}
	k_m \approx \frac{(2m+1)\pi}{2L+1} - \frac{r}{2L+1}\cot\left(\frac{(2m+1)\pi}{4L+2}\right)
\end{eqnarray}
for ground mode $m=0$ and first excited mode $m=1$. The energy becomes correspondingly
\begin{eqnarray}
	\frac{E_m}{t} =-2+\frac{1}{L^2}\left[\frac{(2m+1)^2\pi^2}{4}-2r\right]
\end{eqnarray}

This indicates a crossover at $r=1/L$, i.e., $\epsilon_{i}/t=1+1/L$, as confirmed exactly by the boundary conditions. Indeed, at this point, the ground mode becomes linear,
\begin{eqnarray}
	\psi^0_\ell \propto L+1-\ell
\end{eqnarray}
with energy exactly at $E=-2t$. Meanwhile, the first excited mode is a standing-wave mode with wave vector $k_1$.

\subsubsection{Two mode projection}

In the hypercube limit, the blockaded-chain model can be projected onto two modes along each direction, and achieve generic scaling behaviors. We retain the two lowest exact eigenstates of each finite chain, $|\phi_0\rangle$ and $|\phi_1\rangle$. Near the crossover, their separation is of order $\sim t/L^2$, giving rise to effective dynamics scaling in this order. Without blockade, these two states are exactly the two modes spanning the low energy dynamics.

A graph edge excludes simultaneous occupation of site $\ell=L$ on the two adjacent chains. This constraint leads to an effective overlap between the two modes. To look for an orthonormal basis $|\varphi_0\rangle$ and $|\varphi_1\rangle$, we need a rotation
\begin{eqnarray}
	|\varphi_0\rangle &=& (\gamma|\phi_0\rangle-|\phi_1\rangle)/\sqrt{\gamma^2+1},\\
	|\varphi_1\rangle &=& (|\phi_0\rangle+\gamma|\phi_1\rangle)/\sqrt{\gamma^2+1}
\end{eqnarray}
where $\gamma=\phi^1_{\ell=0}/\phi^0_{\ell=0}\approx\sin(3\pi/(2L+1))/\sin(\pi/(2L+1))$ relates the wavefunction at the edges of the original modes. Notably, this imposes an effective hopping between the two new basis modes,
\begin{eqnarray}
	\tilde{t} = -\langle\varphi_0|h|\varphi_1\rangle =\frac{\gamma}{1+\gamma^2}(E_1-E_0),
\end{eqnarray}
where $E_m$ are the energies of original basis modes $|\phi_m\rangle$. In other words, we have spanned the system onto an effective two-mode basis,
\begin{eqnarray}
	H_\mathrm{eff} = -\frac{\gamma}{1+\gamma^2}(E_1-E_0) \sigma_x + \frac{\gamma^2-1}{2(1+\gamma^2)}(E_1-E_0) \sigma_z \nonumber \\
\end{eqnarray}
The Pauli matrices are spanned within the basis of $|\varphi_m\rangle$.

The projection thus yields the PXP Hamiltonian on the same graph with renormalized hopping strengths. This reveals a minimun gap of the hypercube limit model given by the energy scale of $\Delta_{L,d=L-1} = \tilde{t}/t \Delta_\mathrm{PXP}$, as confirmed numerically by Fig.~\ref{fig:hardness_index}.

\subsubsection{Annealing difficulty for PXP implementation}

\begin{figure}[!t]
	\centering
	\includegraphics[width=0.6\columnwidth]{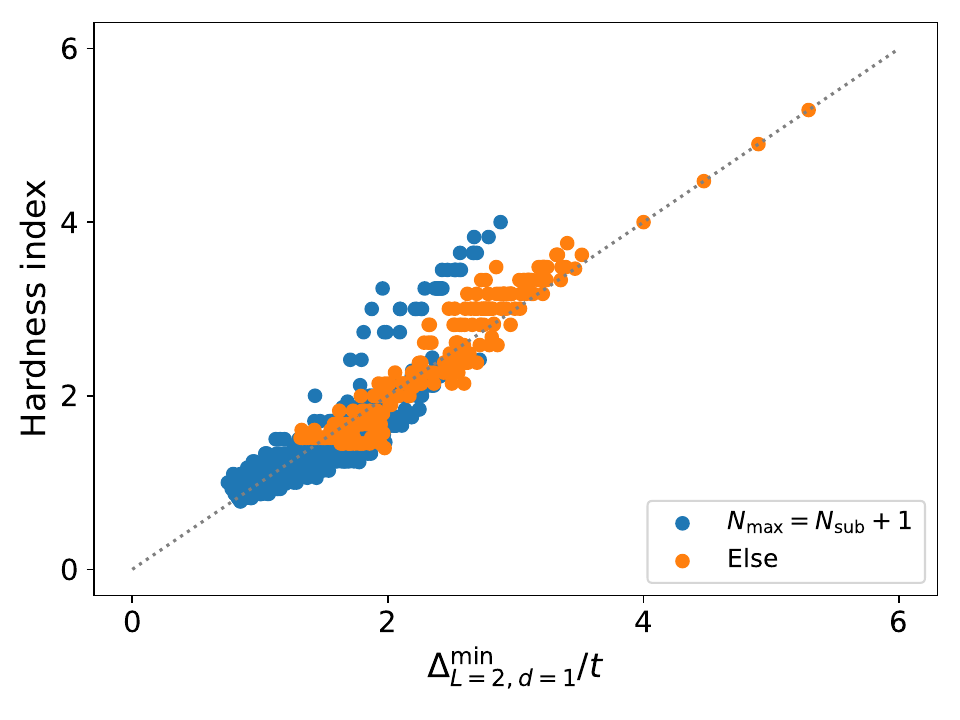}
	\caption{The classical hardness index, Eq.~\eqref{eq:hardness_index_appendix}, compared to the minimum gap along protocol for the PXP implementation, $L=2,d=1$, for graphs with $N\le7$ vertices.
	}
	\label{fig:hardness_index}
\end{figure}

The small-graph spectra also motivate an empirical indicator based on the degeneracies of competing independent-set families. We extract an empirical index,
\begin{eqnarray}
	\label{eq:hardness_index_appendix}
	\mathcal{H} = 
	\begin{cases}
		\dfrac{N_\mathrm{max}+\sqrt{q_\mathrm{max}}-1}{N_\mathrm{sub}+\sqrt{q_\mathrm{sub}}-1}, \quad N_\mathrm{max} = N_\mathrm{sub} + 1 \neq 1 \\
		\\
		2\dfrac{N_\mathrm{max}+\sqrt{q_\mathrm{max}}-1}{{N_\mathrm{max}}^{1.2}}, \quad \mathrm{else}
	\end{cases}
\end{eqnarray}
Here, $N_\mathrm{max/sub}$ refers to the cardinality of the maximum or suboptimal independent set, and $q_\mathrm{max/sub}$ refers to the degeneracy. 
When the cardinality of the maximum and suboptimal independent sets differ by $1$, the quantum annealing performance is mainly determined by the competition between these two independent sets. $N+\sqrt{g}-1$ thus provides a rough estimate to the weight of the independent set, and motivated by the weight $W_q=\sqrt{q}$ of a $q$-clique when $L=2,d=1$. The classical hardness index $\mathcal{H}$ estimates the ratio between the weights of the maximum and suboptimal independent clique sets. When $N_\mathrm{max}-N_\mathrm{sub}>1$, the maximum set does not compete with the suboptimal one, and determines the gap on its own. This index indeed shows a clear correlation with the minimum gap along the annealing in Fig.~\ref{fig:hardness_index}.

\section{Binding in $\mathcal{K}_q$ graphs}
\label{sec:deformation_MICS_appendix}

This Appendix complements the discussion of bound states in complete graphs, $\mathcal{K}_q$, Sec.~\ref{sec:binding_scattering}, and we derive the bound state self-consistency equation, Eq.~\eqref{eq:bound_energy_alpha} in details.

To investigate the effects of coordination enhancement of scattering processes on the bare logical-zero state $|\psi^0\rangle^{\otimes q}$, we split the full Hilbert space into the low-energy subspace spanned solely by $|\psi^0\rangle^{\otimes q}$ and its complement, $Q_q = \mathds{1} - |\psi^0\rangle^{\otimes q}\langle \psi^0|$. This allows us to write down the Feshbach equation,
\begin{eqnarray}
	E - q E_0 = q \tilde{t}^2 \Sigma_q(E).
\end{eqnarray}
Here $qE_0$ is the energy of $|\psi^0\rangle^{\otimes q}$, $\tilde{t} = t\psi^0(d)$ is the effective coupling strength, and the self energy is 
\begin{eqnarray}
	\Sigma_q(E) = \left\langle B \left\vert \frac{1}{E-Q_qHQ_q} \right\vert B \right\rangle.
\end{eqnarray}
with $|B\rangle = Q_q H (\mathds{1}-Q_q)|\psi^0\rangle^{\otimes q}$, or explicitly
\begin{eqnarray}
	|B\rangle = \frac{1}{\sqrt{q}}\sum_{n=1}^{q} |n,d+1\rangle\otimes_{m\neq n}|\psi^0\rangle_m
\end{eqnarray}

To reduce the dimensionality of the problem, we introduce a mapping onto one dimension. Specifically, we consider the variational ansatz generated by a one-dimensional wavefunction $|\varphi\rangle$ through the isometric map $W$,
\begin{eqnarray}
	W|\varphi\rangle = \frac{1}{\sqrt q}
	\sum_{j=1}^q |\varphi\rangle_j \bigotimes_{i\neq j}|\psi^0\rangle_i.
\end{eqnarray}
where we impose $\langle \varphi|\psi^0\rangle=0$. Particularly, the state $|B\rangle$ can be generated using $W$ as $W|d+1\rangle=|B\rangle$.
This mapping, $W$, induces a relation between $Q_q$, and its one-dimensional counterpart, $Q_1 = W^\dagger W = \mathds{1} - |\psi^0\rangle\langle\psi^0|$, as
\begin{eqnarray}
	W^\dagger Q_qHQ_qW = Q_1hQ_1+(q-1)E_0Q_1.
\end{eqnarray}
The variational ansatz is split into contributions from one $|\varphi\rangle$ and $q-1$ logical-zero modes.

Defining $\varepsilon = E - (q-1) E_0$, the Feshbach equation becomes
\begin{eqnarray}
	\varepsilon - E_0 = q \tilde{t}^2 \tilde{\Sigma}_1(\varepsilon),
\end{eqnarray}
where the factor of $q$ comes from the summation in $|B\rangle$ and represents the $q$ different scattering processes, and
\begin{subequations}
\begin{eqnarray}
	\tilde{\Sigma}_1(\epsilon) = 	\left\langle d+1\left| \frac{1}{\varepsilon-Q_1hQ_1} \right|d+1\right\rangle.
\end{eqnarray}
Here, we approximate the complementary-space resolvent by restricting it to the subspace generated by $W$,
\begin{equation}
	W^\dagger(E-Q_qHQ_q)^{-1}W \approx [\varepsilon-Q_1hQ_1]^{-1}.
\end{equation}
This neglects virtual excursions outside the retained subspace.

To proceed to evaluate the self energy, we need to transform it into the bare one-dimensional self energy without projector,
\begin{eqnarray}
	\Sigma_1(\epsilon) = 	\left\langle d+1\left| \frac{1}{\varepsilon-h} \right|d+1\right\rangle.
\end{eqnarray}
\end{subequations}
Splitting the one-dimensional Hamiltonian into $h = E_0|\psi^0\rangle\langle\psi^0| + Q_1 h Q_1$, the Dyson equation relates the bare and projected self energies, $\tilde{\Sigma}_1$ and $\Sigma_1$ as
\begin{eqnarray}
	\Sigma_1^{-1} = \tilde{\Sigma}_1^{-1} - \frac{\tilde{t}^2}{\varepsilon-E_0}
\end{eqnarray}
Substitution leads to
\begin{eqnarray}
	\varepsilon - E_0 = (q-1) \tilde{t}^2 \Sigma_1(\varepsilon).
\end{eqnarray}
Since $1/(\varepsilon-h)$ is the usual one-dimensional propagator, it can be straightforwardly evaluated.

The factor $q$ initially arises from the coherent coupling of the bare logical-zero state $|\psi^0\rangle^{\otimes q}$ to $q$ equivalent scattering channels. After Dyson resummation, virtual processes going through $|\psi^0\rangle^{\otimes q}$ are also taken into account, which reduces the factor from $q$ to $q-1$. However, this reduction is merely algebraic, and should not be interpreted as a reduction in scattering channels.

The energy of the dressed state is directly associated to its density profile, particularly whether it is bound to the logical-zero sector or propagating outwards. Particularly, bound states are characterized energetically by $E<(q-1)E_0-2t$, and their wavefunctions decay with rate $\kappa_q$ along the lattice. Denoting $E-(q-1)E_0 = -2t \cosh\kappa_q$, the propagator for a one-dimensional lattice can be written as 
\begin{eqnarray}
	G = - \frac{1-e^{-2\kappa_q(d+1)}}{2t\sinh\kappa_q}
\end{eqnarray}
where $\kappa_q$ is related to $\alpha_q$ defined in Sec.~\ref{sec:binding_scattering} through $\alpha_q = \kappa_q(d+1)/\pi$. In other words, the wavefunction decays as $\Psi\sim e^{-\pi\alpha_q L/(d+1)}$. This further applies to the minimum gap for $\mathcal{K}_q$.

\section{The recursive generation algorithm}
\label{sec:recursive_algorithm}

The blockaded-chain models are solved typically using exact diagonalization with sparse matrices. This makes the required computational resources formidable as it scales with $L^N$, where $L$ is the length of the lattice, and $N$ is the number of vertices in the graph. In our work, there are two families of graphs with clear patterns, the complete graphs $\mathcal{K}_N$ and the diamond-chain graphs $\mathcal{D}_\nu$. These graphs themselves can be generated recursively. Utilizing the patterns, we can also generate their Hamiltonian and solve their spectra recursively. This method can easily be further generalized for other families of graphs. The algorithm comes in two major steps, mode truncation and iterative generation.

\subsubsection{Mode truncation}

The lattice of the blockaded-chain model has a shape given by Eq.~\eqref{eq:hilbert_space_structure}. It can be split into multiple hypercuboids, and each hypercuboid admits a straightforward solution. 

Formally, this can be treated by the following Hamiltonian,
\begin{eqnarray}
	H = 
	\begin{pmatrix}
		H_A & -t \mathds{1} & -t \mathds{1} & 0 & \ldots \\
		-t \mathds{1} & H_B & 	-t \mathds{1} & 0 & \ldots \\
		-t \mathds{1} & -t \mathds{1} & H_C & 	-t \mathds{1} & \ldots \\
		0 & 0 & -t \mathds{1} & H_D  & \dots \\
		\ldots & \ldots & \ldots & \ldots & \ddots
	\end{pmatrix}
\end{eqnarray}
where $H_{A,B,C,D}$ are Hamiltonians easily diagonalizable describing the sectors A, B, C, D,
\begin{eqnarray}
	H_i = U_i^\dagger \Lambda_i U_i,
\end{eqnarray}
where $\Lambda_i=\mathrm{diag}(\lambda_0,\lambda_1,\dots)$ is a diagonal matrix containing the eigenvalues of $H_i$. Furthermore, $-t\mathds{1}$ indicates a coupling between the two corresponding sectors. Generically for blockaded-chain models, this coupling is takes a uniform value of $t$ when it exists. In this particular example, it couples the sector pairs $(A,B)$, $(A,C)$, $(B,C)$, and $(C,D)$. It is thus possible to construct the overall transform matrix,
\begin{eqnarray}
	U = 
	\begin{pmatrix}
		U_A & 0 & 0 & 0 & \ldots \\
		0 & U_B & 0 & 0 & \ldots \\
		0 & 0 & U_C & 0 & \ldots \\
		0 & 0 & 0 & U_D  & \ldots \\
		\ldots & \ldots & \ldots & \ldots & \ddots
	\end{pmatrix}
\end{eqnarray}
such that
\begin{eqnarray}
	U^\dagger H U = 
	\begin{pmatrix}
		\Lambda_A & -t U_A^\dagger U_B & -t U_A^\dagger U_C & 0 & \ldots \\
		-t U_B^\dagger U_A & \Lambda_B & 	-t U_B^\dagger U_C & 0 & \ldots \\
		-t U_C^\dagger U_A & -t U_C^\dagger U_B & \Lambda_C & 	-t U_C^\dagger U_D & \ldots \\
		0 & 0 & -t U_D^\dagger U_C & \Lambda_D  & \ldots \\
		\ldots & \ldots & \ldots & \ldots & \ddots
	\end{pmatrix} \nonumber \\
\end{eqnarray}
The matrices $(U_i^\dagger U_{i+1})_{mn} = \sum_{n} \psi_{i,m}^* \psi_{i+1,n}$ encode the overlap between the basis wavefunctions from the two neighbouring sub-systems. When the neighboring sectors share the same Hamiltonian $H_i=H_{i+1}$, these matrices are trivially unities $U_i^\dagger U_{i+1} = \mathds{1}$. 

Upon this diagonalization, it is thus possible to truncate the full Hamiltonian $U^\dagger HU$, and drop the high-energy modes with large $\lambda$. This can significantly reduce the computational complexity.

\subsubsection{Motivation from classical algorithm}

\begin{figure}[!t]
	\centering
	\includegraphics[width=\columnwidth]{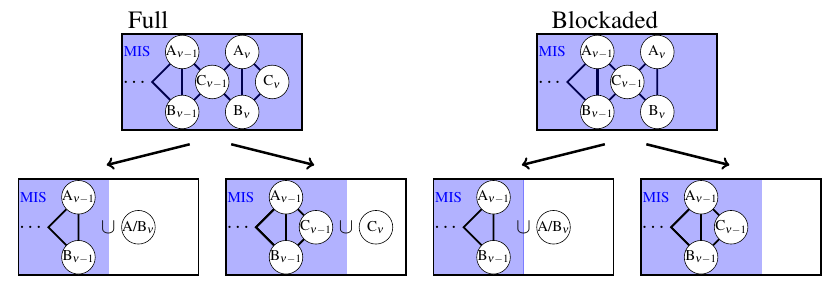}
	\caption{The branching algorithm for classically solving the MIS problem on diamond chain graphs. The MIS solution to the $\nu$-graph is given by the better solution between the two branches, and can thus be generated using the MIS solution to the $\nu-1$-graph.}
	\label{fig:chain_graphs_iter_classic}
\end{figure}

We motivate the quantum algorithm from a classical algorithm for solving the MIS problem on the diamond chain graphs. The \emph{branching algorithm}~\cite{gao2024automated} generates the MIS iteratively, as illustrated in Fig.~\ref{fig:chain_graphs_iter_classic}. Starting with a single vertex as $\nu=0$, in each iteration we attach a $\mathcal{K}_3$ to $\mathcal{D}_{\nu-1}$ for generating $\mathcal{D}_\nu$. The MIS for $\mathcal{D}_{\nu}$ comes from either of the two branches:
\begin{subequations}
\begin{itemize}
	\item Either of the vertices $A_\nu$ or $B_\nu$ belongs to MIS of $\mathcal{D}_\nu$. In this case, $C_{\nu-1}$ cannot belong to the MIS. If we construct an auxiliary blockaded diamond chain graph $\mathcal{D}^\mathrm{block}_{\nu}=\mathcal{D}_{\nu}\setminus\{C_{\nu}\}$, the MIS of $\mathcal{D}_\nu$ can be constructed as the union of the MIS of the blockaded graph $\mathcal{D}^\mathrm{block}_{\nu-1}$ together with $A_\nu$ or $B_\nu$,
	\begin{eqnarray}
		\mathrm{MIS}_1\left(\mathcal{D}_\nu\right) = \mathrm{MIS}
		\left(\mathcal{D}_{\nu-1}\setminus\{C_{\nu-1}\}\right)\cup\{A_\nu\text{ or }B_\nu\}
	\end{eqnarray}

	\item Neither of the vertices $A_\nu$ or $B_\nu$ belongs to MIS of $\mathcal{D}_\nu$. In this case, the MIS $\mathcal{D}_\nu$ can be constructed as the union of the MIS of $\mathcal{D}_{\nu-1}$ together with $C_\nu$,
		\begin{eqnarray}
		\mathrm{MIS}_2\left(\mathcal{D}_\nu\right) = \mathrm{MIS}
		\left(\mathcal{D}_{\nu-1}\}\right)\cup\{C_\nu\}
	\end{eqnarray}
\end{itemize}
The MIS is thus the one with more vertices,
\begin{eqnarray}
	\mathrm{MIS}\left(\mathcal{D}_\nu\right) = \max_{N}\left[\mathrm{MIS}_1\left(\mathcal{D}_\nu\right), \, \mathrm{MIS}_2\left(\mathcal{D}_\nu\right)\right].
\end{eqnarray}
The branching algorithm provides a recursive relation between the MISs of the full and blockaded graphs.
\end{subequations}

\subsubsection{The recursive generation algorithm for the quantum problem}

\begin{figure}[!t]
	\centering
	\includegraphics[width=\columnwidth]{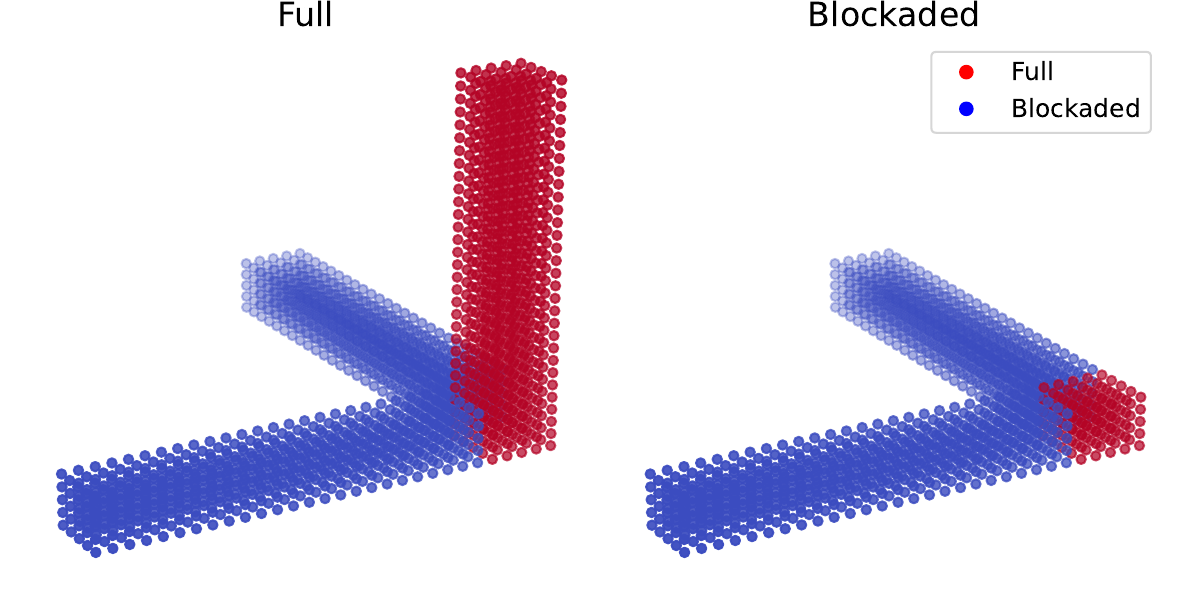}
	\caption{Constructing the effective Hamiltonian for $\nu$-graph from the spectrum of $(\nu-1)$-graph. In each iteration, a \emph{full} and a \emph{blockaded} Hamiltonian are constructed and solved. For both of them, the red (blue) sites host the spectrum of the full (blockaded) Hamiltonian from the last iteration.}
	\label{fig:chain_iterative_illustration}
\end{figure}

A similar strategy can be applied for solving the low-energy spectra of the blockaded chain models based on diamond chain graphs. After introducing an auxiliary blockaded Hamiltonian, an effective Hamiltonian for $\mathcal{D}_\nu$, denoted as $H_\nu$, can be constructed using the spectrum of $H^\mathrm{full}_{\nu-1}$ and $H^\mathrm{block}_{\nu-1}$. The strategy is illustrated in Fig.~\ref{fig:chain_iterative_illustration}. This method allows for truncation for the low-energy spectra, which significantly reduces the computational complexity.

\begin{widetext}
The ``full'' Hamiltonian is defined on lattice sites $\ell^3=(\ell_1,\ell_2,\ell_3)\in\Lambda$ allowed by a fully connected $N=3$-vertex graph. In other words, $1\leq \ell_i \leq L$, and only one of them is allowed to be greater than $L-d$, see Fig.~\ref{fig:chain_iterative_illustration}. The Hamiltonian thus reads
\begin{eqnarray}
	H^\mathrm{full}_{\nu} = 
	 -\sum_{\ell_\alpha^3,m_\alpha,\ell_\beta^3,m_\beta} t^{\nu-1}_{\ell_\alpha^3,m_\alpha,\ell_\beta^3,m_\beta}  |\ell_\alpha^3,m_\alpha\rangle \langle \ell_\beta^3,m_\beta| 
	+ \sum_{\ell^3,m} E^{\nu-1}_{\ell^3,m} |\ell^3,m\rangle \langle \ell^3,m| 
\end{eqnarray}
The effective local energy shift provided by the spectrum of the $(\nu-1)$-graphs is
\begin{eqnarray}
	E^{\nu-1}_{\ell^3,m} = 
	\begin{cases}
		E^{\nu-1, \mathrm{full}}_{m} - \delta, &\quad \ell_{1,2}\leq L-d,\quad \text{(red)}, \\
		E^{\nu-1, \mathrm{blockaded}}_{m} - \delta, &\quad \ell_1> L-d \text{ or } \ell_2> L-d,\quad \text{(blue)}
	\end{cases}
\end{eqnarray}
These two regions are colored in Fig.~\ref{fig:chain_iterative_illustration} by red and blue, respectively.
Here, $E^{\nu-1, \mathrm{full}/\mathrm{blockaded}}_m \in \{\mathrm{eigvalue}(H)^{\nu-1, \mathrm{full}/\mathrm{blockaded}}\}$.
$\delta$ is the chemical potential on boundary condition, $\delta = \#_1 \epsilon_i + \#_L \epsilon_L$, where $\#_{1,L}$ denotes the number of $\ell_i$ being $1$ or $L$, respectively.

Moreover, the effective hopping is provided by the overlap between the eigenmodes of the $(\nu-1)$-graphs, 
\begin{eqnarray}
	t^{\nu-1}_{\ell_\alpha^3,m_\alpha,\ell_\beta^3,m_\beta} = 
	\begin{cases}
		0, &  |\ell_{1/2/3,\alpha} - \ell_{1/2/3,\beta}| \neq 1 \\
		\delta_{m_\alpha,m_\beta}, &\text{not } \ell_{1/2,\alpha} = L-d, \ell_{1/2,\beta} = L-d+1 \\
		\langle \Psi^\mathrm{full}_{m_\alpha} | \Psi^\mathrm{bloc}_{m_\beta} \rangle, &\ell_{1/2,\alpha} = L-d, \ell_{1/2,\beta} = L-d+1
	\end{cases}
\end{eqnarray}
In other words, within the same (blue or red) region, the effective coupling is unity between the same mode of neighboring sites, and is zero in all other cases. Along the boundary between the two regions, the effective coupling is given by the overlap between modes.
\end{widetext}

The ``blockaded'' Hamiltonian can be straightforwardly derived from the full one by blockading the prohibited region,
\begin{eqnarray}
	H^\mathrm{blockaded}_{\nu}  = P_{\ell_3 \leq d} H^\mathrm{full}_{\nu} P_{\ell_3 \leq d}
\end{eqnarray}
where $P_{\ell_3 \leq d}$ is the projection operator onto $\ell_3 \leq d$, cf. Fig.~\ref{fig:chain_iterative_illustration}.

These two Hamiltonians together provide an iterative relation,
\begin{eqnarray}
	\{H^\mathrm{full}_{\nu-1},H^\mathrm{blockaded}_{\nu-1}\} \to \{H^\mathrm{full}_{\nu},H^\mathrm{blockaded}_{\nu}\}.
\end{eqnarray}
By choosing a sufficiently large truncation for the number of modes considered $m=1,2,\dots,M$, we can successively obtain the low-energy spectrum for the implementations of different diamond-chain graphs. In our simulations, we obtain converged results with $M=40$. The same algorithm can be straightforwardly generalized to any other series of graphs of clear pattern. In the main text, it is also applied to the complete graphs $\mathcal{K}_N$.

\subsubsection{Dynamics in the truncated instantaneous basis}
\label{sec:recursive_algorithm_dynamics}

We simulate the real-time dynamics in the subspace spanned by the $M$ lowest instantaneous modes obtained from the recursive algorithm. The quantum state is propagated following an evolution-and-projection strategy. The coefficients between step $t$ and $t+1$ are related through
\begin{eqnarray}
	c_{t+\delta t} =  S^\dagger_{t,t+\delta t} e^{-i \mathcal{E}_t \delta t} c_t.
\end{eqnarray}
Here $\mathcal{E}_t$ is a diagonal matrix containing the instantaneous spectrum at time $t$, associated to the instantaneous modes $|\mu(t)\rangle$. $S_{t,t+\delta t}$ stores the overlaps between the instantaneous modes at time $t$ and $t+\delta t$, ${\left(S_{t,t+\delta t}\right)}_{\mu,\nu} = \langle \mu(t) | \nu(t+\delta t)\rangle$. For each step, this overlap matrix can be computed iteratively from $\nu=1$ to the required $\nu$. Particularly, this numerical dynamical evolution is not exactly unitary due to mode truncation, and norms of the quantum states can decrease during a mode-unconverged evolution.

\end{appendices}

\bibliographystyle{apsrev}
\bibliography{References}

\end{document}